\documentclass[5p,sort,compress,nopreprintline]{elsarticle}  % [], [3p,12pt], [5p,sort&compress]

\usepackage{amssymb}
\usepackage{amsmath}
\usepackage{natbib}
\usepackage{geometry}
\usepackage{graphicx}
\usepackage{txfonts}
\usepackage{hyperref}
\usepackage{microtype}

\usepackage{comment}
\usepackage{multirow}
\usepackage{makecell}
\usepackage{booktabs}
\usepackage{enumitem}
\usepackage{adjustbox}
\usepackage{pbalance}

\usepackage{xcolor}
\newcommand{\revisedA}[1]{{\color{black}#1}}
\newcommand{\revisedB}[1]{{\color{black}#1}}

\newcommand{\CustomizedWidth}{\linewidth}
\makeatletter
\def\fps@figure{t!}
\def\fps@table{t!}
\makeatother
\begin{document}

\begin{frontmatter}

\title{\revisedA{Hybrid Dual-Batch and Cyclic Progressive Learning for Efficient Distributed Training of Convolutional Neural Networks}\tnoteref{aam}}

\tnotetext[aam]{
This is the author accepted manuscript of an article published in \textit{Future Generation Computer Systems}.
The version of record is available at \url{https://doi.org/10.1016/j.future.2026.108892}.
\copyright~2026. This manuscript version is made available under the CC-BY-NC-ND 4.0 license: \url{https://creativecommons.org/licenses/by-nc-nd/4.0/}.
}

\author[1,2]{Kuan-Wei Lu}
\ead{andylu6046@iis.sinica.edu.tw}

\author[2]{Ding-Yong Hong\corref{cor1}}
\ead{dyhong@iis.sinica.edu.tw}

\author[3]{Pangfeng Liu}
\ead{pangfeng@csie.ntu.edu.tw}

\author[2]{Jan-Jan Wu}
\ead{wuj@iis.sinica.edu.tw}

\cortext[cor1]{Corresponding author}

\affiliation[1]{
    organization={Taiwan International Graduate Program in Artificial Intelligence of Things, Academia Sinica and National Taiwan University},
    %addressline={No.~128, Sec.~2, Academia Rd., Nangang Dist.},
    city={Taipei},
    %postcode={115201},
    %state={},
    country={Taiwan}
}

\affiliation[2]{
    organization={Institute of Information Science, Academia Sinica},
    %addressline={No.~128, Sec.~2, Academia Rd., Nangang Dist.},
    city={Taipei},
    %postcode={115201},
    %state={},
    country={Taiwan}
}

\affiliation[3]{
    organization={Department of Computer Science and Information Engineering, National Taiwan University},
    %addressline={No.~1, Sec.~4, Roosevelt Rd., Da'an Dist.},
    city={Taipei},
    %postcode={106216},
    %state={},
    country={Taiwan}
}

%% Abstract, research highlights, and keywords
%% Abstract
\begin{abstract}
Distributed deep learning is critical for training \revisedA{convolutional neural network (CNN)} models on large-scale image datasets with numerous parameters.
Current research primarily focuses on leveraging additional hardware resources and powerful computing units to accelerate the training process.
As a result, larger batch sizes are often employed to speed up training.
However, training with large batch sizes can lead to lower accuracy due to poor generalization.
To address this issue, we propose the \emph{dual-batch learning} scheme, a distributed training method built on the parameter server framework.
This approach maximizes training efficiency by utilizing the largest batch size that the hardware can support while incorporating a smaller batch size to enhance model generalization.
By using two different batch sizes simultaneously, this method improves accuracy with minimal additional training time.
Additionally, to mitigate the time overhead caused by dual-batch learning, we propose the \emph{cyclic progressive learning} scheme.
This technique repeatedly and gradually increases image resolution from low to high during training, thereby reducing training time.
By combining cyclic progressive learning with dual-batch learning, our \emph{hybrid} approach improves both model generalization and training efficiency \revisedA{for image classification tasks}.
%Experimental results using ResNet-18 show that, compared to conventional training methods, our method improves accuracy by 3.3\% while reducing training time by 10.1\% on CIFAR-100, and %improves accuracy by 0.1\% while reducing
%reduces training time by 34.8\% on ImageNet.
Experimental results with ResNet-18 demonstrate that, compared to conventional training methods, our approach improves accuracy by 3.3\% while reducing training time by 10.1\% on CIFAR-100, and further achieves a 34.7\% reduction in training time on ImageNet.
\end{abstract}


\begin{comment}
\begin{highlights}
\item Proposes hybrid dual-batch and cyclic progressive learning for distributed CNNs.
\item Balances generalization and speed using two batch sizes with model-update factors.
\item Introduces cyclic progressive resizing acting as a spectral curriculum for CNNs.
\item Achieves 3.3\% higher accuracy on CIFAR-100 and 34.7\% speedup on ImageNet.
\item Mitigates network congestion and GPU under-utilization in homogeneous clusters.
\end{highlights}
\end{comment}

%% Keywords
\begin{keyword}
convolutional neural networks \sep
image classification \sep
distributed training \sep
parameter server \sep
batch processing \sep
progressive learning
\end{keyword}

\end{frontmatter}


\section{Introduction}

Deep learning has achieved remarkable success in a wide range of applications in recent years.
Notably, it has excelled in computer vision tasks, including image classification~\cite{krizhevsky2017imagenet, simonyan2015very, he2016deep}, object detection~\cite{ren2017faster, liu2016ssd, redmon2016you}, semantic segmentation~\cite{shelhamer2017fully, chen2018deeplab, howard2019searching}, and many others.

Despite these advancements, training deep learning models remains computationally intensive due to the large number of parameters and high workload.
Additionally, hardware architecture imposes inherent limitations on computational capability.
For example, the number of physical cores limits the degree of parallelism, while memory capacity limits the maximum amount of data that can be processed simultaneously.
Consequently, enlarging the batch size while extending the parallelism to reduce training time has become crucial in the training of \revisedA{convolutional neural networks (CNNs)}.

To overcome hardware memory and computing power limitations, researchers have proposed \emph{distributed learning}~\cite{bennun2019demystifying, verbraeken2021a, shi2021a}.
By integrating multiple processors into a training cluster, both computing power and memory capacity can be significantly expanded beyond the limits of a single processor.
Recent studies~\cite{goyal2018accurate, smith2018dont, hu2019a} have demonstrated the effectiveness of distributed training with large-scale GPU and TPU clusters.
Through aggregating computation across numerous processors, batch sizes can be scaled from hundreds to thousands, significantly reducing training time.

% heterogeneous training, hence batch size different
\begin{comment}
However, not all researchers have abundant computing resources and can use many processors to conduct experiments.
%There are studies~\cite{yang2020boa, tyagi2020taming, chen2020semidynamic} focusing on batch tuning for distributed training rather than just increasing the batch size.
Our research does not seek to merely enlarge the batch size as much as possible to reduce training time, as used in conventional training methods.
Instead, we aim to improve model accuracy as much as possible through \emph{batch tuning}~\cite{yang2020boa, tyagi2020taming, chen2020semidynamic} with a slight increase in training time.
The baseline is the training results achieved using the \emph{largest} batch size allowed on all available GPUs.
\end{comment}

However, while training with large batch sizes enhances hardware utilization and accelerates the training process, it often results in reduced model accuracy and poor generalization.
This issue arises because large-batch training reduces gradient diversity, causing models to converge toward sharp minima in the loss landscape~\cite{keskar2017on}.
Such minima correspond to solutions that fit the training data well but generalize poorly to the unseen data.
Moreover, the reduced stochastic noise in large-batch updates weakens the implicit regularization effect that smaller batches can provide, which further contributes to overfitting and decreased generalization ability.

To address the challenges of large-batch training, we propose \emph{dual-batch learning} (DBL), a method that trains CNN models with two batch sizes simultaneously.
We use the maximum batch size supported by the hardware to maximize training efficiency, and incorporate a smaller batch size to enhance gradient diversity and improve generalization.
%Since the two batch sizes have different processing speeds, training data must be proportionally distributed to maximize throughput, and their contributions to the global weight update balanced accordingly.
%To this end, we introduce a contribution adjustment strategy that scales the update ratio based on the amount of data processed by the small and large batches.
Since the two batch sizes have different processing speeds, training data must be proportionally distributed to maintain throughput, and their contributions to the global weight update balanced accordingly.
To this end, we introduce a contribution adjustment strategy that scales the update ratio based on the data processed by the small and large batches.
The preceding methods were published in IEEE COMPSAC 2022~\cite{lu2022efficient}.

\begin{figure*}[t]
    \centering
    \includegraphics[width=.9\linewidth]{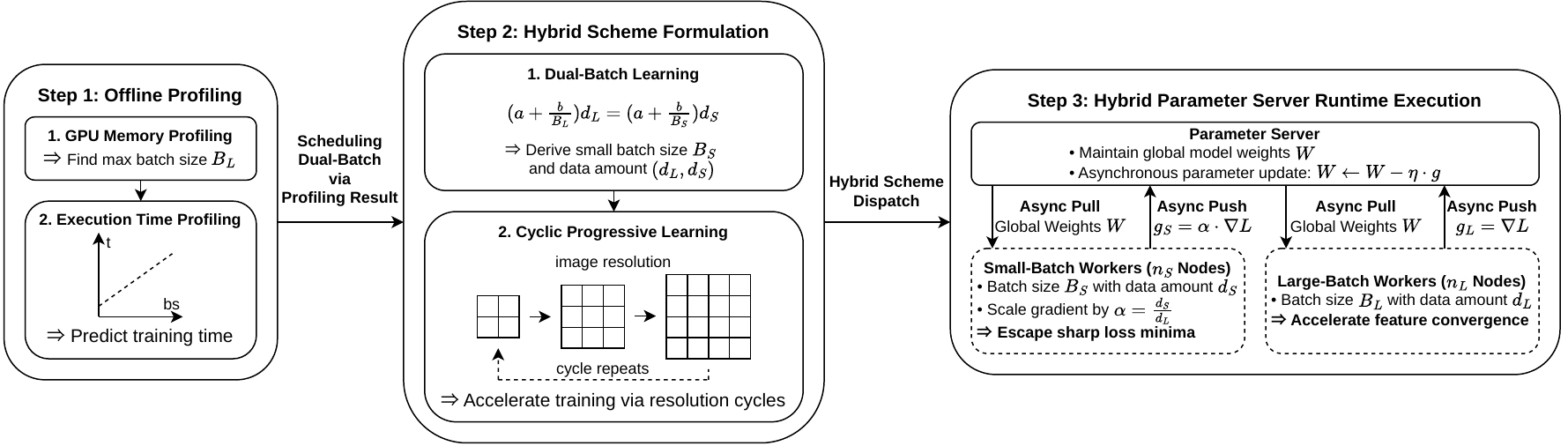}
    \caption{\revisedB{
        Overview of the proposed hybrid distributed training framework.
        Step 1 profiles GPU memory and per-batch latency to formulate the dual-batch allocation.
        Step 2 formulates the hybrid scheme via dual-batch learning and cyclic progressive learning.
        Step 3 coordinates parameter server execution with scaled stochastic gradients.
    }}
    \label{fig:overview}
\end{figure*}

\revisedB{
Building upon our previous work, this work further optimizes the dual-batch learning scheme to reduce overall training time.
To achieve this, we propose \emph{cyclic progressive learning} (CPL), a scheduling strategy that divides the training process into multiple stages and progressively adjusts the image resolutions from low to high within each stage.
By introducing low-resolution images, it reduces computational cost and improves overall training efficiency.
By integrating cyclic progressive learning with dual-batch learning, this \emph{hybrid} approach not only enhances generalization through small batches and exposing models to images of varying resolutions, but also shortens training time through utilizing smaller images in each training stage.
Figure~\ref{fig:overview} illustrates the overall workflow of the proposed distributed framework, including offline profiling, cyclic progressive learning, and dual-batch training.
}

The contributions of this paper are as follows:
\begin{itemize}[nosep]
    \item We propose a novel training strategy that uses two different batch sizes simultaneously to enhance training efficiency and model accuracy \revisedA{for CNN models}.
    Moreover, we introduce the \emph{model-update factor} to balance the contributions of different batch sizes to the global parameters.
    \revisedB{
    \item We present a cyclic progressive learning strategy to reduce training time through an adaptive learning schedule that adjusts image resolutions within training stages.
    \item We integrate the dual-batch scheme and cyclic progressive learning scheme into a hybrid method.
    %\revisedA{for distributed CNN image classification}.
    Experimental results show that our hybrid method reduces training time by 10.1\% while improving accuracy by 3.3\% on CIFAR-100, and reduces training time by 34.7\% while maintaining comparable accuracy on ImageNet.
    }
\end{itemize}

The remainder of this paper is organized as follows.
Section~\ref{sec:background} provides an overview of distributed learning, including the impact of batch size, the training framework, and existing synchronization schemes.
Section~\ref{sec:DBL} explains the dual-batch learning scheme, and
Section~\ref{sec:CPL} introduces cyclic progressive learning and the hybrid scheme.
Section~\ref{sec:experiments} presents the experimental results that evaluate the effectiveness of the proposed method.
Section~\ref{sec:limitations} discusses the limitations of our work and directions for future work.
Finally, Section~\ref{sec:conclusion} concludes the paper.

\section{Background and Related Work}
\label{sec:background}

\subsection{Stochastic Gradient Descent}
\label{sec:sgd}

Stochastic gradient descent (SGD) is a standard optimization algorithm that updates model parameters using gradients.
Equation~\ref{eq:SGD} shows the formulation of SGD:
\begin{equation}
    w_{t+1} = w_t - \eta g = w_t - \eta ( \frac{1}{B} \sum_{x \in X_B} \nabla L(x, w_t) )
    \label{eq:SGD}
\end{equation}
where $w_t$ represents the model weights at iteration $t$, $\eta$ is the learning rate, $X_B$ is a batch of size $B$ sampled from the training dataset, and $L$ is the loss function.

\revisedA{
The batch gradient $\bar{g}$ computed over a sampled subset of data is a sample mean estimator of the true gradient over the entire dataset. According to the \emph{central limit theorem}, the covariance of the batch gradient estimator approximately follows Equation~\ref{eq:grad_variance}:
%is inversely proportional to the batch size $B$:
\begin{equation}
    C(\bar{g}) \approx \frac{1}{B} C(g)
    \label{eq:grad_variance}
\end{equation}
where $C(\bar{g})$ and $C(g)$ represent the covariance of the batch gradient and the single-sample gradient, respectively.
As the batch size $B$ increases, the stochastic noise embedded in the gradient updates decreases.
Consequently, large batch sizes yield more stable updates, whereas small batch sizes introduce greater gradient deviations (stochastic noise) into the parameter optimization trajectory.
}

\subsection{The Effect of Batch Size}
\label{sec:batch_size_effect}

\begin{figure}
    \centering
    \includegraphics[width=\CustomizedWidth]{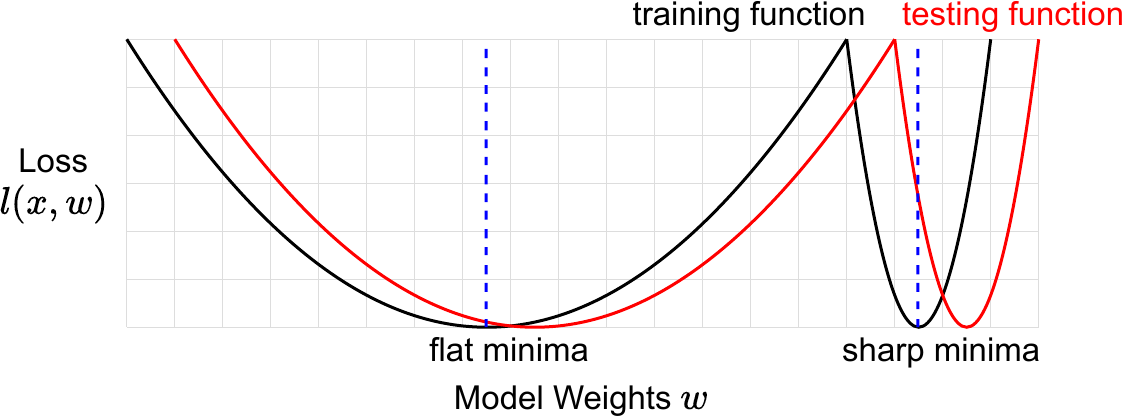}
    \caption{
        Illustration of sharp and flat minima in the loss landscape.
        A sharp minimum leads to higher testing loss and poor generalization, while a flat minimum causes a small increase in loss, thereby improving generalization~\cite{keskar2017on}.
    }
    \label{fig:sharp_flat_minima}
\end{figure}

\paragraph{Sharp and Flat Minima} The level of stochastic noise in gradient updates plays a crucial role in the generalization capability of deep neural networks.
Keskar et al.~\cite{keskar2017on} first demonstrated that the generalization gap, where large-batch training degrades testing performance, correlates closely with the sharpness of the converged minimizers.
They introduced the concepts of sharp and flat minima, as illustrated in Figure~\ref{fig:sharp_flat_minima}.
A \emph{sharp minimum} $\hat{w}$ occurs when the loss increases rapidly near $\hat{w}$, making the model highly sensitive to small parameter perturbations and leading to poor generalization on unseen data.
Conversely, a \emph{flat minimum} $\bar{w}$ exhibits a slowly increasing loss in its neighborhood, which is robust to perturbations and leads to higher test accuracy.

Figure~\ref{fig:sharp_flat_minima} illustrates the loss (y-axis) as a function of model weights (x-axis).
The black line, \emph{training loss function}, indicates the loss computed on the training data.
The red line, \emph{testing loss function}, shows the loss computed using the test data that were not part of the training set.
If the model converges to a sharp minimum on the training loss function, the loss increases significantly when shifting to the testing loss function.
In contrast, if the model converges to a flat minimum, the loss increases only slightly on unseen test data.

\revisedA{
\paragraph{Sharpness of Minima} To quantify the sharpness of a converged minimum, Keskar et al.~\cite{keskar2017on} proposed a sensitivity-based metric.
The sharpness of a minimizer $w$ is characterized by the eigenvalues of the Hessian matrix $\nabla^2 L(w)$, which describe the curvature of the loss landscape~\cite{hochreiter1997flat}.
However, directly computing the Hessian eigenvalues is computationally prohibitive for deep neural networks due to the high dimensionality of the parameter space.
Instead, the sharpness can be approximated by measuring the maximum increase in the loss within a local neighborhood.
Specifically, the perturbation is projected from the $n$-dimensional parameter space onto a lower $p$-dimensional subspace, where $p \ll n$.
To restrict the search space, a constraint box $C_\epsilon$ is defined as:
\begin{equation}
    C_\epsilon = \{ z \in \mathbb{R}^p : -\epsilon(|(A^g w)_i| + 1) \le z_i \le \epsilon(|(A^g w)_i| + 1) \}
    \label{eq:sharpness_box}
\end{equation}
where $A \in \mathbb{R}^{n \times p}$ is a random projection matrix, $A^g$ is the pseudo-inverse of $A$, and $\epsilon$ controls the neighborhood size.
This box defines the boundaries of the neighborhood and scales the allowed perturbation according to the magnitude of each parameter component.
The sharpness metric, denoted by $\phi_{w,L}(\epsilon, A)$, is then calculated as the worst-case relative change of the loss function:
\begin{equation}
    \phi_{w,L}(\epsilon, A) = \frac{(\max_{y \in C_\epsilon} L(w + Ay)) - L(w)}{1 + L(w)} %\times 100 % x100 represents a percentage and is not needed
    \label{eq:sharpness_metric}
\end{equation}
where $y \in \mathbb{R}^p$ represents the perturbation vector within the subspace, and $Ay \in \mathbb{R}^n$ maps this perturbation back to the full parameter space.
By finding the worst-case loss increment in a low-dimensional neighborhood, this metric characterizes the sharpness of the converged minima.
}

\revisedA{
%\paragraph{Optimization Trajectories with Batch Sizes}
\paragraph{Optimization Trajectories}
The optimization trajectories under different batch sizes explain why training converges to different types of minima:
\begin{itemize}[nosep]
    \item \textbf{Small-Batch Training ($B_S$)}:
    According to Equation~\ref{eq:grad_variance}, a small batch size introduces high gradient variance, causing stochastic perturbations in the optimization trajectory.
    These perturbations help the optimizer to escape the narrow basins of sharp minima.
    %, which are characterized by large positive eigenvalues in the Hessian $\nabla^2 L(w)$.
    As a result, the optimizer explores a broader region of the loss landscape and is more likely to converge to flatter minima with better generalization.
    \item \textbf{Large-Batch Training ($B_L$)}:
    A large batch size substantially reduces gradient variance, leading to a smoother and more deterministic optimization trajectory.
    As a result, the reduced stochasticity makes it more difficult for the optimizer to escape sharp minima during training, increasing the likelihood of convergence to sharper solutions.
    %As a result, the optimization process tends to remain in a local minimum close to the initial weights.
\end{itemize}

Takase et al.~\cite{takase2018why} further analyzed this behavior from the perspective of loss landscape stability.
They observed that large-batch training produces relatively stable gradient updates, thereby changing the loss surface only slightly between iterations.
%keeping the geometry almost identical between iterations and preventing the parameters from escaping suboptimal local basins.
%In contrast, small batch sizes introduce instability (shifting landscape) that dynamically alters the loss geometry between updates, facilitating escape from poor minima.
In contrast, small-batch training produces higher stochasticity, continuously perturbing the loss landscape.
This behavior resembles the exploration mechanism of simulated annealing, where strong noise in the early optimization stages encourages broader exploration before converging to a favorable solution.

These insights highlight a fundamental trade-off between training efficiency and generalization.
Large-batch training offers higher computational efficiency but is more susceptible to sharp minima, while small-batch training improves exploration and favors flatter minima at the cost of lower hardware utilization and longer training time.
This trade-off motivates our proposed dual-batch learning scheme, which combines the complementary advantages of training with both batch sizes to balance optimization efficiency and generalization performance.
%larger batch sizes are computationally efficient but trap the optimizer in sharp minima, while smaller batch sizes facilitate escape and convergence to flat minimizers at the cost of training throughput.
%This motivates the need to dynamically adjust or combine batch configurations during training to simultaneously balance exploration (generalization accuracy) and exploitation (convergence speed), which forms the core justification for our proposed dual-batch learning scheme.
}

\subsection{Parameter Server Framework}

The parameter server framework~\cite{smola2010an, power2010piccolo, ahmed2012scalable, li2014scaling, li2014communication} is a widely used method for centralized distributed training.
\revisedB{A parameter server system consists of a logical server and multiple workers.}
The server maintains the global model, updates parameters, and coordinates the workers.
The worker is responsible for the training computation.
A worker receives parameters from the server, trains the model, and sends the results back to the server.

In a parameter server framework, model training proceeds as follows.
At the start of each iteration, each worker requests (or \emph{pulls}) the global model from the server.
The workers then replace their local models with the received global model.
Next, the workers train the model and update their parameters locally.
Once training is complete, the workers send (or \emph{push}) the updated parameters back to the server, which then updates the global model.
This cycle repeats until the model converges or reaches a predetermined number of iterations.

\subsection{Synchronization Strategy}

Prior works have proposed several synchronization strategies to achieve both data consistency and high performance in distributed learning across multiple machines~\cite{gerbessiotis1994direct, chen2017revisiting, recht2011hogwild, dean2012large, chilimbi2014project, ho2013more, cui2014exploiting, zhang2018stay, zhao2019dynamic}.
These strategies can be broadly classified into three categories: bulk synchronous parallel (BSP), asynchronous parallel (ASP), and stale synchronous parallel (SSP).

The \emph{bulk synchronous parallel} (BSP)~\cite{gerbessiotis1994direct, chen2017revisiting} is a commonly used synchronization scheme that operates with a \emph{barrier}.
At the start of each iteration, the server informs workers to pull the global model.
Each worker then trains the model and computes gradients independently.
After training, the worker pushes the updated parameters back to the server and waits for synchronization.
Once the server receives parameters from all workers, it updates the global model and notifies all workers to begin the next iteration.
The advantage of BSP is that all workers have the same local model at the start of each iteration, ensuring consistency.
However, the disadvantage is that all workers must wait for the slowest one to push its parameters, which can result in longer training time.

The \emph{asynchronous parallel} (ASP)~\cite{recht2011hogwild, dean2012large, chilimbi2014project} is a synchronization scheme that operates  without a barrier.
Unlike BSP, where the server waits for all workers to push their gradients before proceeding, each worker in an ASP model \emph{independently} pulls the global model from the server and updates the model without waiting for other workers.
Thus, when a worker computes gradients and pushes them back, the server immediately updates the model without synchronization.
This results in shorter training time since there is no waiting.
However, the inconsistency between local models can lead to difficulties in convergence.

The \emph{stale synchronous parallel} (SSP)~\cite{ho2013more, cui2014exploiting} is a flexible synchronization scheme that lies between ASP and BSP.
SSP introduces a hyper-parameter, the \emph{staleness threshold} $s$, which limits the iteration gap between the fastest and slowest workers.
When the gap is within $s$, the workers behave like an ASP model, independently communicating with the server.
If the gap reaches $s$, the server suspends the fastest worker until the slowest worker catches up and updates its local parameters.
SSP becomes ASP when $s$ is infinite, allowing workers to proceed without synchronization, and SSP becomes BSP when $s$ is 0, requiring workers to synchronize at every iteration.

\revisedA{
\subsection{Progressive Learning}

The design of progressive learning for CNNs has evolved through several key developments.
Howard~\cite{howard2018training} first proposed \emph{progressive resizing}, which trains models on low-resolution images during early epochs and gradually increases the input resolution.
The key intuition is that coarse, low-resolution images are sufficient for learning fundamental visual features, while higher-resolution images enable the model to capture finer details.
%This strategy reduces training time but often suffers from a degradation in final accuracy.
Touvron et al.~\cite{touvron2019fixing} addressed the discrepancy between training and testing resolutions, showing that training on lower-resolution images while testing on higher-resolution images reduces computational cost and improves final validation accuracy.
Hoffer et al.~\cite{hoffer2019mix} extended progressive learning by dynamically sampling multiple image resolutions per batch, enhancing scale invariance and accelerating model convergence.
However, these methods employ a fixed regularization strategy throughout training, regardless of the input resolutions.

Tan and Le~\cite{tan2021efficientnetv2} observed that using fixed regularization during progressive learning causes accuracy degradation.
They identified that strong regularization on low-resolution images tends to cause underfitting, whereas weak regularization on high-resolution images leads to overfitting.
To address this issue, they proposed an \emph{adaptive regularization} method that dynamically adjusts dropout ratios and data augmentation strength according to the image resolution, allowing progressive learning to scale efficiently without sacrificing generalization.
}
\section{Dual-Batch Learning}
\label{sec:DBL}

\subsection{Overview}

We propose the \emph{dual-batch learning} scheme that employs two batch sizes simultaneously during model training on a parameter server, where some workers use small batches while others use large batches.
This approach reduces test loss and thereby improves model accuracy.
The two batch sizes are designed to combine the stability of gradient descent associated with large batches and the strong generalization capability provided by small batches.
Furthermore, we adopt \emph{asynchronous parallel} training to accommodate the different training speeds resulting from the two batch sizes.
Using a synchronous scheme would cause the smaller batch size to slow down the overall training process, thus reducing training efficiency.

The dual-batch learning scheme addresses three key challenges.
First, it predicts the training time associated with different batch sizes to allocate input data appropriately.
Second, it determines optimal batch sizes and distributes the data accordingly to achieve balanced training times among workers.
Finally, it evaluates the relative contribution of each batch size and integrates these contributions into the global model during the weight update phase.

\revisedA{
Note that this work focuses on homogeneous distributed training environments, where all worker nodes have identical GPU capabilities and network bandwidth.
Under this assumption, the execution time of each training batch is highly deterministic.
Therefore, a simple profiling-based linear regression model is sufficient to accurately estimate the training time
%(the accuracy of the proposed cost model is validated in Section~\ref{sec:exp_DBL}).
(with the relative errors evaluated in Section~\ref{sec:exp_DBL}).
Supporting heterogeneous environments requires sophisticated hardware performance models that capture differences in computational capability, memory capacity, and communication bandwidth across devices, resulting in higher modeling complexity and overhead.
Thus, extending the proposed framework to heterogeneous clusters is left for future work.
}

\subsection{Training Time Prediction}
\label{sec:train_time_prediction}

The dual-batch learning scheme requires an accurate prediction of training time as a function of the batch size.
Once the training time can be predicted, this model enables the selection of appropriate batch sizes for the training process.
Thus, an accurate training time prediction model is essential.

The training time of a batch is determined by the amount of training data per batch and the GPU computing speed.
Additionally, the total training time is the product of the training time per batch and the total number of batches.
We assume the training time is a linear function of the batch size $x$, denoted as $ax + b$.
Let the total amount of training data be $d$; the number of batches required for training would then be $\lceil \frac{d}{x} \rceil$.
Thus, the total training time $t$ is given by Equation~\ref{eq:train_time}.
\begin{equation}
    t = (ax+b) \cdot \lceil \frac{d}{x} \rceil      \label{eq:train_time}
\end{equation}
We conduct experiments to verify the correctness of Equation~\ref{eq:train_time}.
For this purpose, we use two widely used machine learning frameworks: PyTorch~\cite{paszke2019pytorch} and TensorFlow~\cite{abadi2016tensorflow}.
The training data contain 50,000 images of size 32$\times$32.
All experiments are performed on an NVIDIA GTX 1080 GPU using a ResNet-18 model.
The model is trained with batch sizes ranging from 1 to 500.
%The training time for one epoch under different batch sizes is in Figure~\ref{fig:train_time_an_epoch}.
%The experiments show that the training time decreases when the batch size increases since large batch sizes utilize the GPU resources more efficiently.

\begin{figure}
    \centering
    \includegraphics[width=\CustomizedWidth]{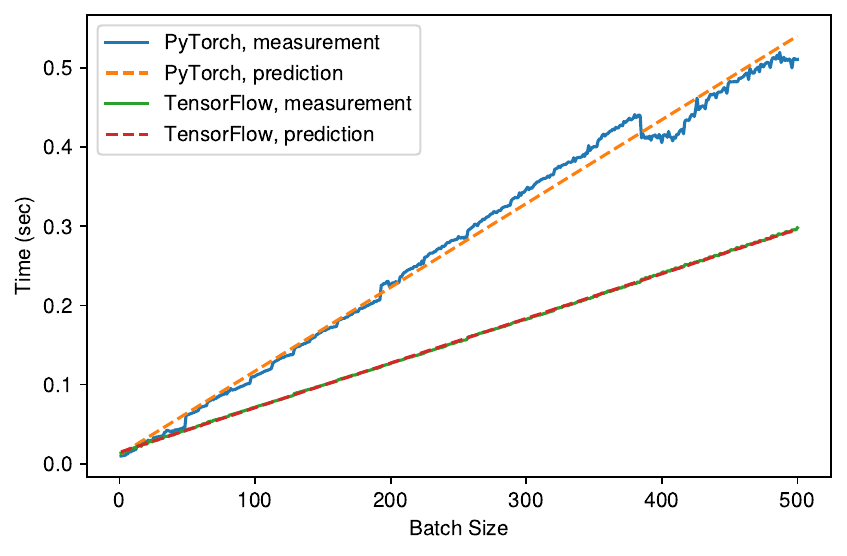}
    \caption{
        Training time per batch for different batch sizes in PyTorch and TensorFlow.
        The results highlight the linear relationship between batch size and training time.
    }
    \label{fig:train_time_a_batch}
\end{figure}

We first validate the assumption that the training time per batch is a linear function of the batch size.
Figure~\ref{fig:train_time_a_batch} presents the training time per batch for different batch sizes in both PyTorch and TensorFlow.
The results confirm a clear linear relationship between batch sizes and training times, allowing us to estimate coefficients $a$ and $b$ via linear regression.
In addition, TensorFlow exhibits more predictable training times than PyTorch.
The variability observed in PyTorch may stem from its back-end implementation, where the framework checks the batch size and invokes different NVIDIA cuDNN~\cite{chetlur2014cudnn} kernel libraries accordingly.
In contrast, TensorFlow is oblivious to the batch size and determines the back-end implementation based on operation parameters and hardware characteristics.
For this reason, we adopt TensorFlow in subsequent experiments to ensure greater predictability.

\begin{figure}
    \centering
    \includegraphics[width=\CustomizedWidth]{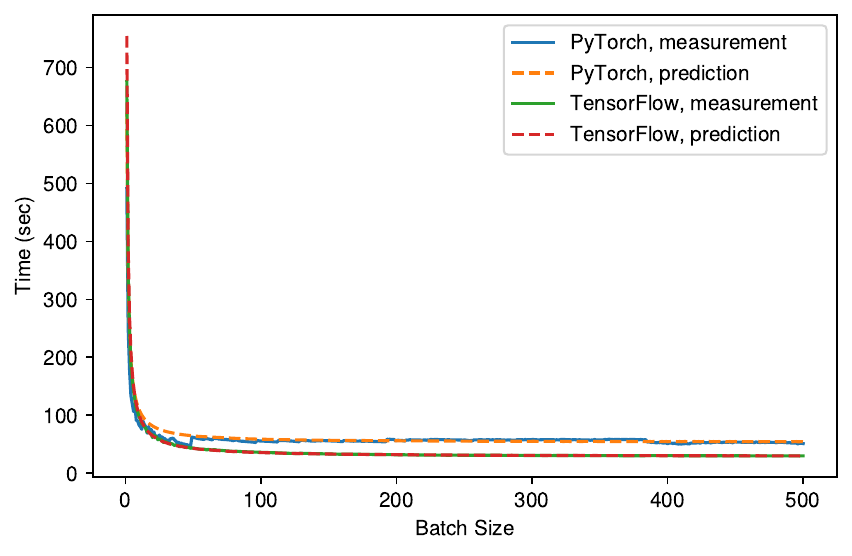}
    \caption{
        Training time per epoch for different batch sizes.
        Larger batch sizes improve GPU utilization and reduce training time, as validated by experimental results.
    }
    \label{fig:train_time_an_epoch}
\end{figure}

Figure~\ref{fig:train_time_an_epoch} presents both the predicted and actual training times per epoch in PyTorch and TensorFlow under various batch sizes.
The results show that training time decreases as the batch size increases, since larger batches make more efficient use of GPU resources.
Moreover, Equation~\ref{eq:train_time} accurately predicts the training time for TensorFlow across all batch sizes.

\begin{figure}
    \centering
    \includegraphics[width=\CustomizedWidth]{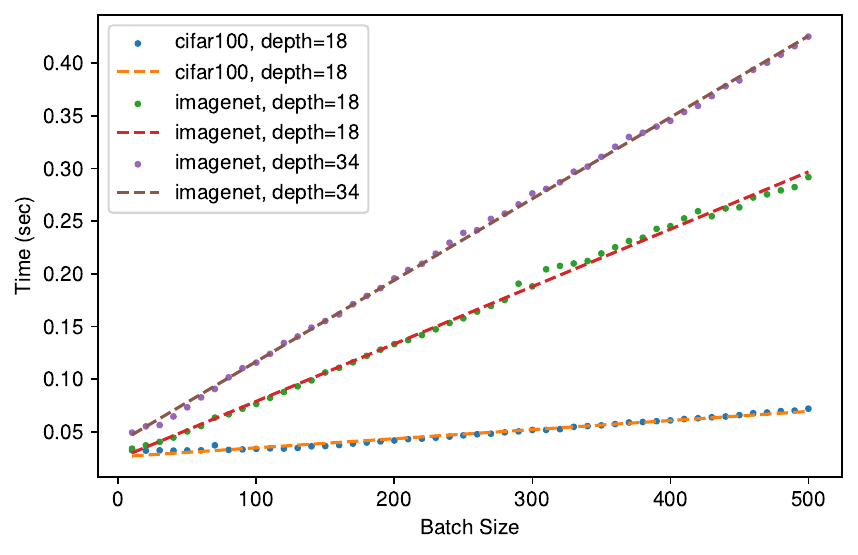}
    \caption{Training time per batch across different model depths and datasets on an RTX 3090, sampled at batch size intervals of 10.}
    \label{fig:linear_generality}
\end{figure}

\revisedB{%Although the linear model is verified on a GTX 1080, we further test whether it holds for other network depths and workloads.
%For this purpose, we measure the training time per batch on an RTX 3090.
To test whether the linear model generalizes across different network depths and workloads, we further measure the performance on an RTX 3090 GPU.
Figure~\ref{fig:linear_generality} illustrates the training time per batch for ResNet-18 on CIFAR-100, ResNet-18 on ImageNet, and ResNet-34 on ImageNet, with batch sizes ranging from 10 to 500 in increments of 10.
%sampled at batch size intervals of 10 up to 500.
%The results demonstrate that the per-batch training time strictly follows a linear trend across all evaluated cases.
The results demonstrate a consistent linear relationship between batch size and training time across all evaluated configurations, confirming the generalizability of the linear model.
Furthermore, since coefficients $a$ and $b$ depend on the computational complexity of models and the underlying hardware, deploying the approach on a different GPU architecture or neural network requires re-profiling.
However, the profiling procedure is lightweight, requiring only a few measurements across several batch sizes before training begins.}

\subsection{Batch Size and Data Amount}

\begin{comment}
The dual-batch learning scheme employs two batch sizes, denoted by $B_S$ (small batch size) and $B_L$ (large batch size), where $B_S < B_L$.
When using $B_L$ on all GPUs, the training time $t$ will be minimized.
However, if we reallocate some GPUs with $B_S$ instead of $B_L$, the training time increases to $\hat{t}$.
In the following, we allocate $\hat{t} = kt > t$ time for training and call $k > 1$ the \emph{extra training time ratio}.
We aim to explore the trade-offs by introducing this extra training time ratio.
\end{comment}
The dual-batch learning scheme employs two batch sizes, denoted by $B_S$ (small batch size) and $B_L$ (large batch size), where $B_S < B_L$.
When all GPUs use $B_L$, the training time per iteration is denoted by $t$, which represents the minimum possible training time.
However, if some GPUs are reassigned to use $B_S$ instead of $B_L$, the training time increases to $\hat{t}$.
In this case, we define $\hat{t} = kt$ with $k > 1$, where $k$ is referred to as the \emph{extra training time ratio}.
This ratio quantifies the additional training time introduced by incorporating smaller batch sizes, enabling us to explore the trade-off between training efficiency and model generalization.

Load balancing is an essential issue in dual-batch learning.
If the same amount of data is allocated to each worker regardless of batch size, the training time will vary across batch sizes, leading to the \emph{straggler problem}.
In this situation, slower workers require significantly longer time to complete their tasks, thus delaying the overall training process.

To address this, we allocate $d_S$ data to each worker with batch size $B_S$ and $d_L$ data to each worker with batch size $B_L$.
%,  ensuring that the workload is evenly distributed across all workers.
%We define the number of large-batch workers $n_L$ and small-batch workers $n_S$, as well as specify the maximum batch size $B_L$ and the total data amount $d$.
Let $n_L$ and $n_S$ denote the numbers of large-batch and small-batch workers, respectively, with $n = n_S + n_L$ representing the total number of workers, and $d$ the total amount of training data.
%Next, our approach sequentially determines the data allocation for each worker, denoted as $d_L$ for large batch workers and $d_S$ for small batch workers, and finally determines the corresponding small batch size $B_S$.
We first simplify Equation~\ref{eq:train_time} into Equation~\ref{eq:train_time_simplified}:
\begin{equation}
    t \approx (a + \frac{b}{x}) \cdot d     \label{eq:train_time_simplified}
\end{equation}
%We then calculate the training time when using only the large batch size $B_L$ on all $n$ workers.
%Given that the total data amount is $d$, the data assigned to each worker is $\frac{d}{n}$.
%According to Equation~\ref{eq:train_time_simplified}, the minimum training time can be expressed as $(a + \frac{b}{B_L}) \frac{d}{n}$.
According to Equation~\ref{eq:train_time_simplified}, when using only the large batch size $B_L$ on all $n$ workers, the minimum training time can be expressed as $(a + \frac{b}{B_L}) \frac{d}{n}$, where $n = n_L$.

%Suppose we use $n_L$ workers with the large batch size $B_L$ and $n_S$ workers with the small batch size $B_S$.
Suppose we use $n_L$ large-batch and $n_S$ small-batch workers.
To ensure a balanced workload between the two types of workers, the training time for both large and small batches should be $k (a + \frac{b}{B_L}) \frac{d}{n}$, representing a duration $k$ times longer than that of using only the large batch size $B_L$ on all $n$ workers, as shown in Equations~\ref{eq:sametime1} and \ref{eq:sametime2}.
\begin{align}
    k (a + \frac{b}{B_L}) \frac{d}{n}   &= (a + \frac{b}{B_L}) d_L  \label{eq:sametime1} \\
                                        &= (a + \frac{b}{B_S}) d_S  \label{eq:sametime2}
\end{align}

From Equation~\ref{eq:sametime1}, we can easily derive that $d_L = k \cdot \frac{d}{n}$.
Additionally, $d_S$ can be obtained from Equation~\ref{eq:find_dS} once $d_L$ is determined.
\begin{equation}
    d = n_L d_L + n_S d_S       \label{eq:find_dS}
\end{equation}
By combining Equations~\ref{eq:sametime1} and \ref{eq:sametime2}, we derive Equation~\ref{eq:sametime3}.
Through transposing terms, Equation~\ref{eq:sametime3} can be transformed into Equation~\ref{eq:find_BS}.
\begin{gather}
    (a + \frac{b}{B_L}) d_L = (a + \frac{b}{B_S}) d_S       \label{eq:sametime3} \\
    B_S = \frac{b}{(a + \frac{b}{B_L}) \frac{d_L}{d_S} - a} \label{eq:find_BS}
\end{gather}

Finally, since $B_L$, $d_L$, and $d_S$ are given, we can use Equation~\ref{eq:find_BS} to determine $B_S$.
The parameters $a$ and $b$ are obtained from Section~\ref{sec:train_time_prediction} via linear regression, and the optimal $B_S$ varies depending on the hardware configuration.

\subsection{Model-Update Factor}

Since different workers process different amounts of data, the importance of each worker's contribution should be different.
Workers using the large batch size $B_L$ handle a more significant amount of data $d_L$, and thus, their parameters are more critical.
In contrast, workers using the small batch size $B_S$ process less data $d_S$, and their parameters are less critical.

During the model update phase, we adjust the contributions of workers with different batch sizes into the global parameters, taking into account the amounts of data processed.
The model-update factor for large-batch workers (batch size $B_L$) is set to $1$.
For small-batch workers (batch size $B_S$), we propose two schemes for the model-update factor.
The first scheme uses a factor of ${d_S}/{d_L}$, which directly reflects the ratio of data processed by the small-batch and large-batch workers.
The second scheme adopts a factor of $\sqrt{{d_S}/{d_L}}$.
%, accounting for the relationship between batch size and gradient variation.
According to the discussion of the batch size effect on the gradient in Section~\ref{sec:batch_size_effect}, the amount of data is proportional to the batch size, which in turn influences the variance of the gradient.
The standard deviation of the gradient, which is the square root of the variance, should reflect the importance of weights updated by the small-batch workers.
A similar argument can be found in the work of Smith et al.~\cite{smith2018dont}.
In summary, when a worker updates the global model, its contribution is scaled by the model-update factor, which depends on the amount of data processed by the worker.
Both schemes ensure balanced contributions between the small- and large-batch workers.
\begin{comment}
Let the model-update factor for workers using $B_L$ be 1.
We define two model-update factors for workers using $B_S$.
The first model-update factor is ${d_S}/{d_L}$, which reflects the amount of data the worker can process.
The second model-update factor is $\sqrt{{d_S}/{d_L}}$.
According to the batch size effect on the gradient discussed in Section~\ref{sec:batch_size_effect}, the amount of data is proportional to the batch size, which in turn affects the variation of the gradient.
The standard deviation of the gradient, which is the square root of the variance, should reflect the importance of the weights updated by the small-batch workers.
A similar argument can be found in the work of Smith et al.~\cite{smith2018dont}.
To summarize, when a worker updates the global model, its contribution is multiplied by the model-update factor, which depends on the amount of data the worker processes.
\end{comment}

\section{Hybrid Dual-Batch and Cyclic Progressive Learning}
\label{sec:CPL}

\subsection{Cyclic Progressive Learning}

In the preceding section, we introduce the extra training time ratio $k$, which represents the additional training time ratio incurred by incorporating small-batch workers to improve model generalization.
While a larger $k$ can enhance generalization through increased gradient diversity, it also extends the overall training duration.
To alleviate this trade-off, we propose \emph{cyclic progressive learning}, which is an improved variant of progressive learning~\cite{tan2021efficientnetv2} and is designed to further reduce training time while maintaining or even enhancing generalization performance in dual-batch learning.
%To further reduce training time and enhance the model's generalization ability in dual-batch learning, we propose the \emph{cyclic progressive learning} scheme, which is an improved version of progressive learning~\cite{tan2021efficientnetv2}.
%This scheme provides two advantages based on varying image resolutions: (1) reduced training time when using low-resolution images, and (2) enhanced accuracy when using high-resolution images.
%Additionally, exposing the model to a broader range of image resolutions helps mitigate overfitting, thereby improving model generalization.

\revisedA{
The motivation for cyclic progressive learning stems from two lines of prior works: \emph{progressive resizing}~\cite{howard2018training} and \emph{adaptive regularization}~\cite{tan2021efficientnetv2}.
Although these progressive learning strategies provide clear advantages, two critical bottlenecks limit their efficiency.
First, existing progressive learning schemes typically employ a \emph{fixed} batch size throughout the training process.
The batch size is determined based on the memory requirement of high-resolution images to avoid GPU memory overflow during the final training phase.
%To avoid GPU memory overflow during the final high-resolution training phase, the batch size must be capped based on the hardware memory limit at the maximum resolution.
Therefore, this causes significant GPU compute and memory under-utilization during the early low-resolution phases, leading to inefficient resource usage and prolonged training time.
Second, because progressive learning monotonically increases resolution over time, high-resolution images are only introduced in the late stages of training, by which time the learning rate has decayed to a very small value.
Although high-resolution images contain rich, fine-grained visual details, the decayed learning rate severely restricts the magnitude of parameter updates (Equation~\ref{eq:SGD}), preventing the network from effectively learning these detailed features.

To overcome these issues, we propose \emph{cyclic progressive learning} (CPL).
Instead of a single monotonic resolution scale, CPL divides the training process into multiple stages, and within each stage, the resolution is cyclically increased from low to high.
This is combined with two key innovations: adaptive batch sizes and learning rate scheduling.
To maximize GPU utilization, we dynamically adjust batch sizes based on image sizes, employing larger batches during low-resolution phases to maximize throughput.
Furthermore, by repeating the resolution cycle within each training stage, high-resolution images are revisited with more favorable optimization conditions, ensuring effective parameter updates with appropriate learning rates for all resolutions.
}

Table~\ref{tab:scheme_comparison} presents an example of the cyclic progressive learning scheme (third row).
In this example, the training process is divided into two stages, each associated with distinct learning rates $\eta_1$ and $\eta_2$.
%named \emph{step} and \emph{stage} respectively.
%For each step, we decrease the learning rate $\eta$ and split the training process into multiple stages.
Within each stage, the image resolution $r$ and dropout rate $d$ are increased progressively every few epochs, while the batch sizes $B$ are dynamically adjusted in accordance with the corresponding image resolutions.
%In addition, we use \emph{dropout}~\cite{hinton2012improving, srivastava2014dropout} as the training regularization method, which randomly discards the training result and sets the weight value to zero according to a user-defined probability $d$.

\begin{table}
    \centering
    \caption{
        Comparison of the proposed training schemes with two stages and two sub-stages.
        The main differences include learning rate $\eta$, batch size $B$, image resolution $r$, and dropout rate $d$.
    }
    \label{tab:scheme_comparison}
    \adjustbox{max width=\linewidth}{
        \begin{tabular}{cccccc}
    \toprule
    \multirow{2}{*}{scheme} & stage & \multicolumn{2}{c}{1} & \multicolumn{2}{c}{2} \\
    \cmidrule(lr){3-4} \cmidrule(lr){5-6}
    & sub-stage & 1 & 2 & 1 & 2 \\
    \midrule
    \multirow{4}{*}{\makecell{dual-batch\\learning}} & LR & \multicolumn{2}{c}{$\eta_1$} & \multicolumn{2}{c}{$\eta_2$} \\
    & BS & \multicolumn{4}{c}{$(B_{S_2}, B_{L_2})$} \\
    & resolution & \multicolumn{4}{c}{$r_2$} \\
    & dropout & \multicolumn{4}{c}{$d_2$} \\
    \midrule
    \multirow{4}{*}{\makecell{cyclic\\progressive\\learning}} & LR & \multicolumn{2}{c}{$\eta_1$} & \multicolumn{2}{c}{$\eta_2$} \\
    & BS & $B_1$ & $B_2$ & $B_1$ & $B_2$ \\
    & resolution & $r_1$ & $r_2$ & $r_1$ & $r_2$ \\
    & dropout & $d_1$ & $d_2$ & $d_1$ & $d_2$ \\
    \midrule
    \multirow{4}{*}{\makecell{hybrid \\scheme}} & LR & \multicolumn{2}{c}{$\eta_1$} & \multicolumn{2}{c}{$\eta_2$} \\
    & BS & $(B_{S_1}, B_{L_1})$ & $(B_{S_2}, B_{L_2})$ & $(B_{S_1}, B_{L_1})$ & $(B_{S_2}, B_{L_2})$ \\
    & resolution & $r_1$ & $r_2$ & $r_1$ & $r_2$ \\
    & dropout & $d_1$ & $d_2$ & $d_1$ & $d_2$ \\
    \bottomrule
\end{tabular}

\begin{comment}
\begin{tabular}{cc|c|c|c|c}
    \toprule
    \multirow{2}{*}{scheme} & stage & \multicolumn{2}{c|}{1} & \multicolumn{2}{c}{2} \\
    & sub-stage & 1 & 2 & 1 & 2 \\
    \midrule
    \multirow{4}{*}{\makecell{dual-batch\\learning}} & LR & \multicolumn{2}{c|}{$\eta_1$} & \multicolumn{2}{c}{$\eta_2$} \\
    & BS & \multicolumn{4}{c}{$(B_{S_2}, B_{L_2})$} \\
    & resolution & \multicolumn{4}{c}{$r_2$} \\
    & dropout & \multicolumn{4}{c}{$d_2$} \\
    \midrule
    \multirow{4}{*}{\makecell{cyclic\\progressive\\learning}} & LR & \multicolumn{2}{c|}{$\eta_1$} & \multicolumn{2}{c}{$\eta_2$} \\
    & BS & $B_1$ & $B_2$ & $B_1$ & $B_2$ \\
    & resolution & $r_1$ & $r_2$ & $r_1$ & $r_2$ \\
    & dropout & $d_1$ & $d_2$ & $d_1$ & $d_2$ \\
    \midrule
    \multirow{4}{*}{\makecell{hybrid \\scheme}} & LR & \multicolumn{2}{c|}{$\eta_1$} & \multicolumn{2}{c}{$\eta_2$} \\
    & BS & $(B_{S_1}, B_{L_1})$ & $(B_{S_2}, B_{L_2})$ & $(B_{S_1}, B_{L_1})$ & $(B_{S_2}, B_{L_2})$ \\
    & resolution & $r_1$ & $r_2$ & $r_1$ & $r_2$ \\
    & dropout & $d_1$ & $d_2$ & $d_1$ & $d_2$ \\
    \bottomrule
\end{tabular}
\end{comment}

    }
\end{table}

\revisedA{
From the perspective of representation learning, CPL regularizes CNN training by controlling the frequency spectrum of the input images.
Low-resolution images retain primary low-frequency spatial structures (e.g., global shapes, outlines, and contours) while filtering out high-frequency noise.
Training on lower resolutions in the early cycles forces the early layers of the CNN to learn robust, low-frequency structural features.
As the resolution cyclically increases, the network progressively integrates high-frequency detailed features (e.g., fine textures and boundaries) onto the established global structures.
This cycle acts as a spectral curriculum that prevents the network from overfitting to high-frequency noise,
%in the early phases of optimization,
thus regularizing the training process and enhancing the final test generalization capability.
}

\subsection{Hybrid Scheme}

Finally, we integrate cyclic progressive learning with the dual-batch learning scheme, resulting in the \emph{hybrid scheme}.
Although dual-batch learning extends the overall training duration, the incorporation of cyclic progressive learning compensates for this by progressively resizing image resolutions, thereby improving training efficiency even when using smaller batches.
Moreover, by combining multiple batch sizes with varying image resolutions, the hybrid scheme further enhances the model's generalization capability.

An example of the hybrid scheme is presented in the fourth row of Table~\ref{tab:scheme_comparison}.
Compared with dual-batch learning alone, the hybrid scheme trains the model in each stage using dynamically adjusted image resolutions, with batch sizes automatically adapted to match these resolutions.
Unlike approaches that employ only cyclic progressive learning, the hybrid scheme extends the batch size configuration within each stage from a single batch size to two distinct batch sizes, effectively leveraging the complementary benefits of both strategies.

\revisedB{
Figure~\ref{fig:overview} illustrates the overall workflow of the proposed hybrid scheme.
Prior to training, offline profiling is conducted to determine the optimal batch configuration and balanced data amount for each target resolution.
During training, the scheduler executes cyclic progressive learning across successive stages, periodically adjusting image resolutions and regularization strengths.
Worker nodes are partitioned into large-batch workers that maximize GPU throughput and small-batch workers that inject stochastic exploration.
The parameter server coordinates asynchronous parameter updates, scaling small-batch gradients by the model-update factor $\alpha = d_S / d_L$ to maintain unbiased optimization.
This integrated scheme allows lower-resolution images to accelerate training while preserving the generalization benefits of dual-batch gradient diversity.
}
\section{Experiments}
\label{sec:experiments}

In this section, we first evaluate the effectiveness of dual-batch learning.
Next, we assess the performance of combining the dual-batch learning scheme with cyclic progressive learning.
Then, we analyze the accuracy of the method that automatically determines the maximum batch size for a given GPU.

\subsection{Training with Dual-Batch Learning}
\label{sec:exp_DBL}

\subsubsection{Experimental Setup}

We use the CIFAR-100~\cite{krizhevsky2009learning} dataset for the image classification task.
CIFAR-100 consists of 50,000 training images and 10,000 testing 32$\times$32 color images in 100 classes, with each class containing 500 training images and 100 testing images.
ResNet-18~\cite{he2016deep} serves as our evaluation model.
This deep neural network consists of eighteen convolutional layers and a fully connected layer.
ResNet-18 uses residual learning to enhance model depth and improve accuracy.
We train ResNet-18 with TensorFlow~\cite{abadi2016tensorflow} version 2.6 on the parameter server framework.
The parameter server framework consists of one server and four workers, forming a GPU cluster.
Each machine is equipped with an Intel Xeon E5-1630 v3 CPU, 64GB RAM, and an NVIDIA GeForce GTX 1080 GPU, running Ubuntu 18.04 LTS Linux.
The devices are connected by 1Gbps Ethernet.

\revisedA{
\subsubsection{Impact of Batch Size on Generalization}

To evaluate the impact of batch sizes on model generalization and convergence, we train ResNet-18 on the CIFAR-100 dataset using a range of batch sizes ($B$) from 32 to 1024.
To isolate the effect of batch size from communication overhead, this experiment is conducted on a single GPU.
The model is trained for 95 epochs, consisting of a 5-epoch linear warm-up phase (the learning rate is increased gradually to 0.1), followed by 90 epochs of training with a cosine learning rate decay schedule.
Table~\ref{tab:impact_bs} lists the final test loss and test accuracy for each batch size, and Figure~\ref{fig:impact_bs} shows the training curves.

\begin{table}
    \centering
    \caption{Impact of batch size on final testing loss and accuracy on CIFAR-100.}
    \label{tab:impact_bs}
    \adjustbox{max width=\linewidth}{
        \begin{tabular}{c|cccccc}
    \toprule
    Batch Size ($B$) & 32 & 64 & 128 & 256 & 512 & 1024 \\
    \midrule
    Testing Loss & 1.823 & 1.807 & 1.760 & 1.748 & 1.914 & 2.136 \\
    Testing Accuracy & 73.7\% & 72.9\% & 71.7\% & 71.5\% & 69.1\% & 67.4\% \\
    \bottomrule
\end{tabular}

    }
\end{table}

\begin{figure}
    \centering
    \includegraphics[width=\CustomizedWidth]{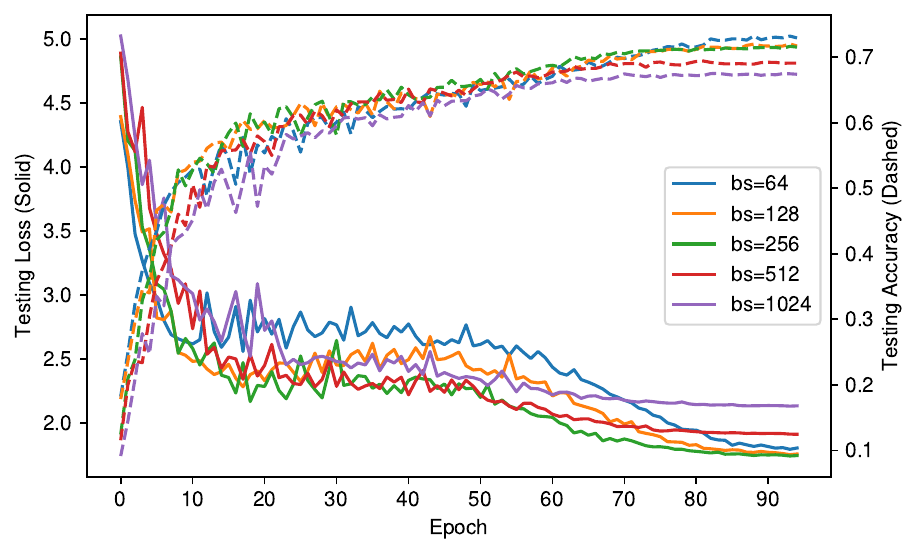}
    \caption{Testing accuracy and loss trajectories for different batch sizes.}
    \label{fig:impact_bs}
\end{figure}

As shown in the results, the test accuracy consistently degrades as the batch size increases.
The accuracy decreases from 73.7\% with $B=32$ to 67.4\% with $B=1024$, while the test loss increases from 1.823 to 2.136.
This observation aligns with the findings of Keskar et al.~\cite{keskar2017on} and Takase et al.~\cite{takase2018why}, proving that large-batch training compromises model accuracy.
This motivates our proposed dual-batch learning scheme, which leverages the speed benefits of large-batch training while incorporating small-batch workers to enhance generalization capability.
}

\subsubsection{Evaluation of Model-Update Factor}

Since small-batch workers train on a smaller amount of data than large-batch workers, it seems reasonable that the training results from small-batch workers should have \emph{less} influence when updating the global model.
To verify this, we compare the accuracy of using the two model-update factors (${d_S}/{d_L}$ and $\sqrt{{d_S}/{d_L}}$) with the case where no model-update factor is applied.
In this experiment, we set the extra training time ratio $k$ to $1.1$.
The lower part of Table~\ref{tab:merge_setting} lists the training configurations, including batch sizes and data amounts for the workers.

\begin{table}
    \centering
    \caption{Training parameters for different numbers of small-batch workers, under extra training time ratios of $1.05$ and $1.1$.}
    \label{tab:merge_setting}
    \adjustbox{max width=\linewidth}{
        % Table: 9 x 7
\begin{tabular}{ccccccc}
    \toprule
    $k$ & $(n_S, n_L)$ & $B_S$ & $d_S$ & $B_L$ & $d_L$ & $\frac{d_S}{d_L}$ \\
    \midrule
    \multirow{4}{*}{1.05} & (1, 3) & 83 & 10625 & \multirow{3}{*}{500} & \multirow{3}{*}{13125} & 0.810 \\
    & (2, 2) & 154 & 11875 &&& 0.905 \\
    & (3, 1) & 205 & 12291 &&& 0.936 \\
    & (4, 0) & 242 & 12500 & - & - & - \\
    \midrule
    \multirow{4}{*}{1.1} & (1, 3) & 38 & 8750 & \multirow{3}{*}{500} & \multirow{3}{*}{13750} & 0.636 \\
    & (2, 2) & 87 & 11250 &&& 0.818 \\
    & (3, 1) & 127 & 12083 &&& 0.879 \\
    & (4, 0) & 160 & 12500 & - & - & - \\
    \bottomrule
\end{tabular}

\begin{comment}
%105
\begin{tabular}{c|c|c|c|c|c}
    \toprule
    $n_S$ / $n_L$ & $B_S$ & $d_S$ & $B_L$ & $d_L$ & $\frac{d_S}{d_L}$ \\
    \midrule
    1 / 3 & 83 & 10625 & \multirow{3}{*}{500} & \multirow{3}{*}{13125} & 0.810 \\
    2 / 2 & 154 & 11875 &&& 0.905 \\
    3 / 1 & 205 & 12291 &&& 0.936 \\
    4 / 0 & 242 & 12500 & - & - & - \\
    \bottomrule
\end{tabular}
%110
\begin{tabular}{c|c|c|c|c|c}
    \toprule
    $n_S$ / $n_L$ & $B_S$ & $d_S$ & $B_L$ & $d_L$ & $\frac{d_S}{d_L}$ \\
    \midrule
    1 / 3 & 38 & 8750 & \multirow{3}{*}{500} & \multirow{3}{*}{13750} & 0.636 \\
    2 / 2 & 87  & 11250 &&& 0.818 \\
    3 / 1 & 127 & 12083 &&& 0.879 \\
    4 / 0 & 160 & 12500 & - & - & - \\
    \bottomrule
\end{tabular}
\end{comment}

    }
\end{table}

\begin{figure}
    \centering
    \includegraphics[width=\CustomizedWidth]{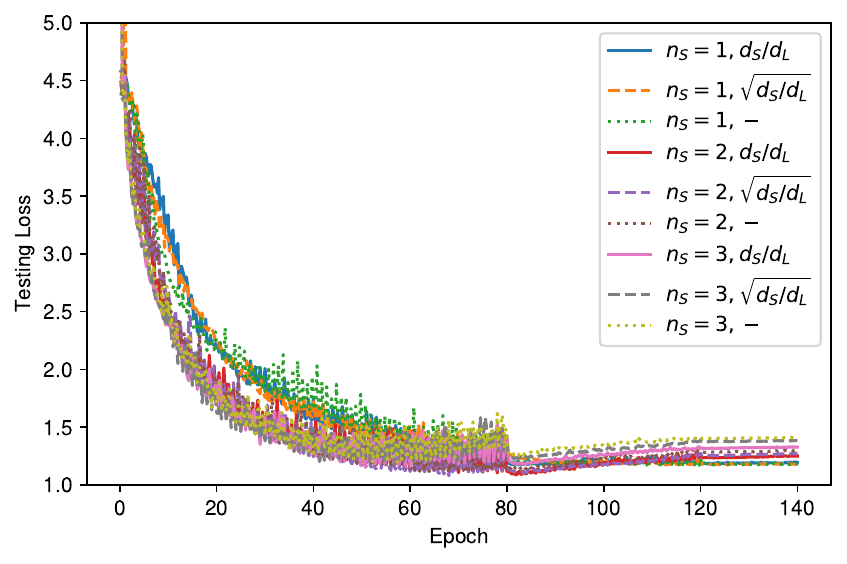}
    \caption{
        Comparison of testing loss under three conditions: ${d_S}/{d_L}$, $\sqrt{{d_S}/{d_L}}$, and without a model-update factor.
    }
    \label{fig:loss_MUfactor}
\end{figure}

\begin{figure}
    \centering
    \includegraphics[width=\CustomizedWidth]{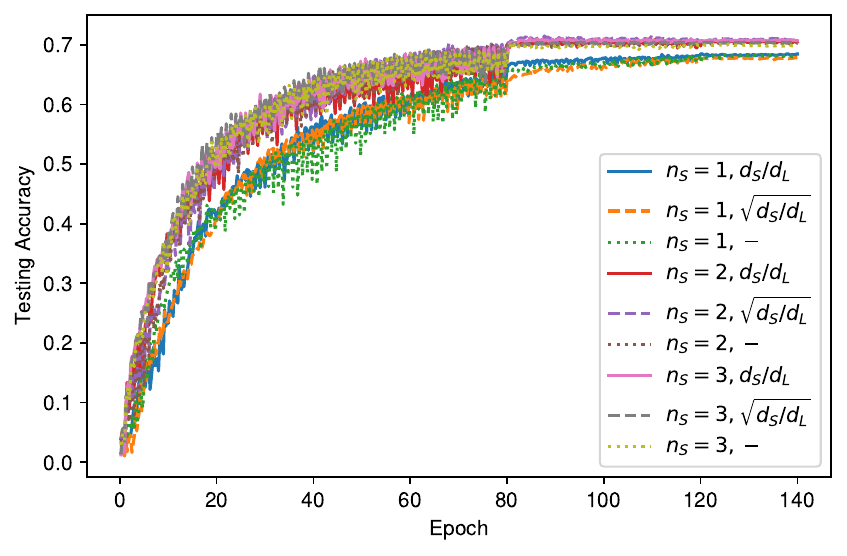}
    \caption{
        Comparison of testing accuracy under three conditions: ${d_S}/{d_L}$, $\sqrt{{d_S}/{d_L}}$, and without a model-update factor.
    }
    \label{fig:acc_MUfactor}
    \end{figure}

Table~\ref{tab:result_MUfactor} presents the performance results, and Figures~\ref{fig:loss_MUfactor} and \ref{fig:acc_MUfactor} respectively illustrate the testing loss and testing accuracy throughout the training process.
The results indicate that using the model-update factor ${d_S}/{d_L}$ improves accuracy in all cases, achieving up to 0.9\% improvement compared to training without it.
In contrast, the model-update factor $\sqrt{{d_S}/{d_L}}$ does not perform as well as ${d_S}/{d_L}$.
Although $\sqrt{{d_S}/{d_L}}$ yields a 0.4\% improvement over ${d_S}/{d_L}$ when training with two small-batch workers, it performs worse in the other two cases.
Since ${d_S}/{d_L}$ consistently outperforms training without the model-update factor in the experiments, we conclude that applying the model-update factor enhances training effectiveness.
Therefore, we will use the model-update factor ${d_S}/{d_L}$ in the remaining experiments.

\begin{table}
    \centering
    \caption{
        Impact of the model-update factor on training results.
    }
    \label{tab:result_MUfactor}
    \adjustbox{max width=\linewidth}{
        \begin{tabular}{cccccccc}
    \toprule
    $n_S$ / $n_L$ & \makecell{model-update\\factor} & value & $B_S$ & $d_S$ & \makecell{testing\\loss} & \makecell{testing\\accuracy} & \makecell{accuracy\\improvement} \\
    \midrule
    \multirow{3}{*}{1 / 3} & $d_S / d_L$ & 0.636 & \multirow{3}{*}{38} & \multirow{3}{*}{8750} & 1.193 & 68.5\% & 0.5\% \\
    & $\sqrt{d_S / d_L}$ & 0.797 &&& 1.179 & 67.8\% & -0.2\% \\
    & - & - &&& 1.187 & 68.0\% & - \\
    \midrule
    \multirow{3}{*}{2 / 2} & $d_S / d_L$ & 0.818 & \multirow{3}{*}{87} & \multirow{3}{*}{11250} & 1.250 & 70.5\% & 0.3\% \\
    & $\sqrt{d_S / d_L}$ & 0.904 &&& 1.270 & 70.9\% & 0.7\% \\
    & - & - &&& 1.295 & 70.2\% & - \\
    \midrule
    \multirow{3}{*}{3 / 1} & $d_S / d_L$ & 0.879 & \multirow{3}{*}{127} & \multirow{3}{*}{12083} & 1.327 & 70.7\% & 0.9\% \\
    & $\sqrt{d_S / d_L}$ & 0.938 &&& 1.383 & 70.4\% & 0.6\% \\
    & - & - &&& 1.411 & 69.8\% & - \\
    \bottomrule
\end{tabular}

    }
\end{table}

\subsubsection{Evaluation of Numbers of Small-batch Workers}

Despite the fact that small-batch workers can achieve lower loss, their number must be sufficient to improve the final accuracy.
According to Equation~\ref{eq:find_BS}, reducing the number of small-batch workers ($n_S$) decreases the optimal small-batch size ($B_S$).
In addition, according to Equation~\ref{eq:find_dS}, the amount of data allocated to these workers ($d_S$) also decreases as $n_S$ decreases.
However, when the number of small-batch workers is too small, they train on much less data than large-batch workers.
As a result, insufficient small-batch training data leads to only minor accuracy improvement.

We conduct experiments to demonstrate the importance of having a sufficient number of small-batch workers.
In these experiments, we set the large-batch size ($B_L$) as high as possible to maximize GPU utilization and reduce training time.
Specifically, we set $B_L$ to 500, which is close to the maximum batch size allowed on the GPU, as larger values cause TensorFlow to trigger memory warnings.
Hence, our baseline for comparison is the configuration where all workers train the model with a batch size of 500.
We use four workers to train the model for 140 epochs.
The initial learning rate is 0.1, and it decreases by a factor of 5 at the 80th and 120th epochs.

For comparison purposes, the baseline method is enhanced using the gradual warm-up strategy~\cite{goyal2018accurate} to improve accuracy.
We start from a learning rate of $\frac{0.1}{5}$ and increase it until it reaches 0.1 after five epochs.
The baseline converges at a loss of 1.429 and an accuracy of 67.8\%.

\paragraph{\underline{$\mathbf{k=1.05}$}}
We first evaluate the performance when $k$ is 1.05.
We set $B_L$ to 500, which is the same as the baseline's batch size.
Given that CIFAR-100 has 50,000 training images, we can derive $d_L$ as 13,125 according to Equation~\ref{eq:sametime1}.
We vary the number of small-batch workers ($n_S$) from 1 to 4.
Correspondingly, the number of large-batch workers ($n_L$) changes from 3 to 0.
Using Equations~\ref{eq:find_dS} and \ref{eq:find_BS}, we compute $d_S$ and $B_S$ for each combination of $n_S$ and $n_L$, since we have determined $d_L$ and $B_L$.
In addition, we also calculate the model-update factor ($\frac{d_S}{d_L}$).
The upper part of Table~\ref{tab:merge_setting} lists these parameters for the training process.

Accurate prediction is essential for the effectiveness of our dual-batch learning.
Hence, we evaluate how accurately Equation~\ref{eq:train_time} can predict the training time as a function of the batch size and corresponding data amount.
We measure the training time by averaging five runs and compare it with the predicted time from Equations~\ref{eq:sametime1} and \ref{eq:sametime2}.
The upper part of Table~\ref{tab:time_merge} shows the measured training time, predicted time, and the relative error across different batch sizes and data amounts.
We observe that the maximum relative error is +3.5\%, occurring at the batch size of 500 and the corresponding data amount of 13,125.
This indicates that our prediction is accurate enough for the dual-batch learning to be feasible.

\begin{table}
    \centering
    \caption{The measured time, predicted time, and relative error for different numbers of small-batch workers, under extra training time ratios of $1.05$ and $1.1$.}
    \label{tab:time_merge}
    \adjustbox{max width=\linewidth}{
        % Table: 11 x 6
\begin{tabular}{cccccc}
    \toprule
    \makecell{extra training\\time ratio} & \makecell{batch\\size} & \makecell{data\\amount} & \makecell{measured\\time} & \makecell{predicted\\time} & \makecell{relative\\error} \\
    \midrule
    \multirow{5}{*}{1.05} & 500 & 13125 & 8.098 & 7.821 & +3.5\% \\
    \cmidrule{2-6}
    & 83 & 10625 & 7.989 & 7.985 & +0.1\% \\
    & 154 & 11875 & 7.615 & 7.816 & -2.6\% \\
    & 205 & 12291 & 7.622 & 7.792 & -2.2\% \\
    & 242 & 12500 & 7.610 & 7.757 & -1.9\% \\
    \midrule
    \multirow{5}{*}{1.1} & 500 & 13750 & 8.182 & 8.193 & -0.1\% \\
    \cmidrule{2-6}
    & 38 & 8750 & 8.183 & 8.398 & -2.5\% \\
    & 87 & 11250 & 8.141 & 8.338 & -2.4\% \\
    & 127 & 12083 & 8.220 & 8.163 & +0.7\% \\
    & 160 & 12500 & 8.222 & 8.135 & +1.0\% \\
    \bottomrule
\end{tabular}

\begin{comment}
%105
\begin{tabular}{c|c|c|c|c}
    \toprule
    \makecell{batch\\size} & \makecell{data\\amount} & \makecell{measured\\time} & \makecell{predicted\\time} & \makecell{relative\\error} \\
    \midrule
    500 & 13125 & 8.098 & 7.821 & +3.5\% \\
    \midrule
    83 & 10625 & 7.989 & 7.985 & +0.1\% \\
    154 & 11875 & 7.615 & 7.816 & -2.6\% \\
    205 & 12291 & 7.622 & 7.792 & -2.2\% \\
    242 & 12500 & 7.610 & 7.757 & -1.9\% \\
    \bottomrule
\end{tabular}
%110
\begin{tabular}{c|c|c|c|c}
    \toprule
    \makecell{batch\\size} & \makecell{data\\amount} & \makecell{measured\\time} & \makecell{predicted\\time} & \makecell{relative\\error} \\
    \midrule
    500 & 13750 & 8.182 & 8.193 & -0.1\% \\
    \midrule
    38 & 8750 & 8.183 & 8.398 & -2.5\% \\
    87 & 11250 & 8.141 & 8.338 & -2.4\% \\
    127 & 12083 & 8.220 & 8.163 & +0.7\% \\
    160 & 12500 & 8.222 & 8.135 & +1.0\% \\
    \bottomrule
\end{tabular}
\end{comment}

    }
\end{table}

\begin{figure}
    \centering
    \includegraphics[width=\CustomizedWidth]{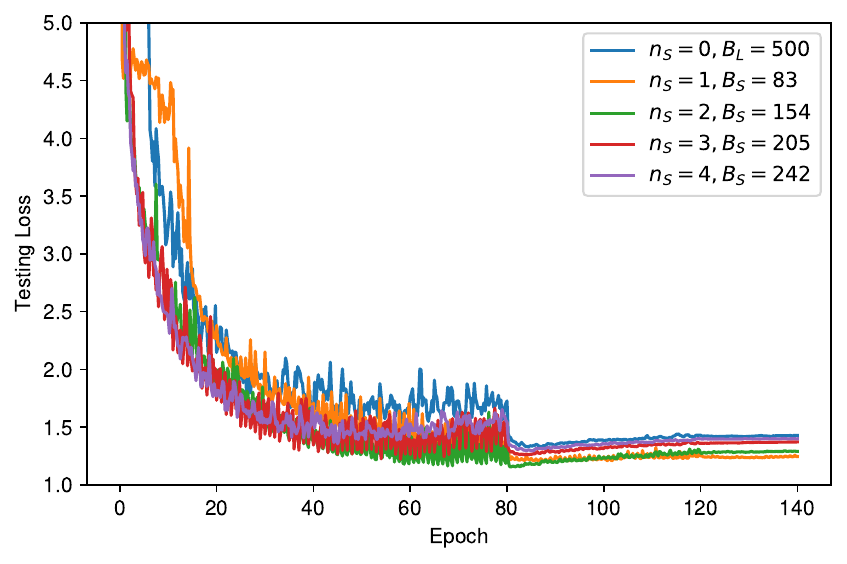}
    \caption{
        Effect of different numbers of small-batch workers on testing loss when $k = 1.05$.
    }
    \label{fig:loss_105}
\end{figure}

\begin{figure}
    \centering
    \includegraphics[width=\CustomizedWidth]{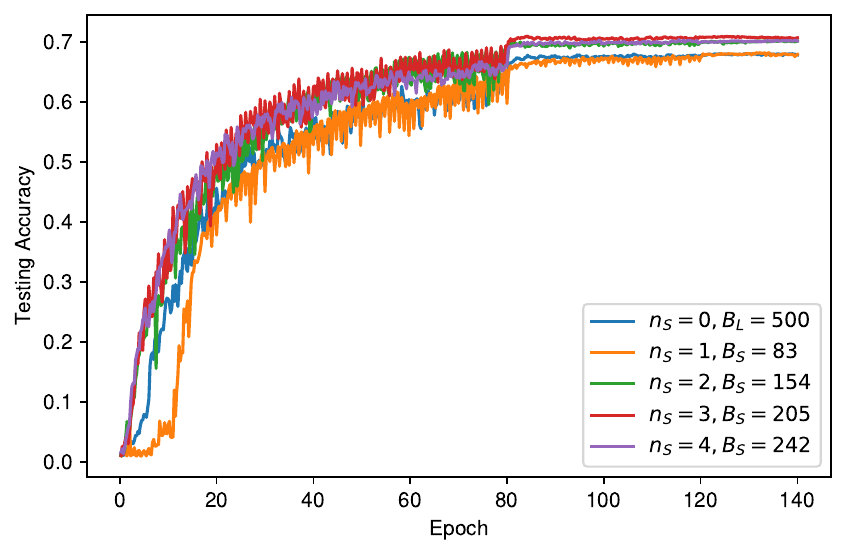}
    \caption{
        Effect of different numbers of small-batch workers on testing accuracy when $k = 1.05$.
    }
    \label{fig:acc_105}
\end{figure}

After verifying the prediction accuracy of training time for different batch sizes, we proceed with training using different numbers of small-batch workers.
The third row of Table~\ref{tab:merge_result} lists the final testing loss and accuracy for all combinations of batch sizes for $k = 1.05$; Figures~\ref{fig:loss_105} and \ref{fig:acc_105} show the testing loss and accuracy of the training process.
We observe that when \emph{one} small-batch worker and \emph{three} large-batch workers are used, we achieve the lowest loss (1.246) among all combinations, and it is 0.183 lower than the baseline method.
However, the accuracy of using only \emph{one} small-batch worker is only 0.1\% better than the baseline.
In contrast, using \emph{three} small-batch workers and \emph{one} large-batch worker results in a higher loss (1.372), which is worse than using just one small-batch worker.
Nevertheless, it achieves the highest accuracy of 70.6\%, 2.8\% higher than the baseline, and has the best accuracy among all combinations.

\begin{table}
    \centering
    \caption{Accuracy results for different numbers of small-batch workers, under extra training time ratios of $1.05$ and $1.1$.}
    \label{tab:merge_result}
    \adjustbox{max width=\linewidth}{
        % Table: 10 x 6
\begin{tabular}{cccccc}
    \toprule
    \makecell{$B_L = 500$\\4 workers} & $(n_S, n_L)$ & $B_S$ & loss & accuracy & \makecell{accuracy\\improvement} \\
    \midrule
    $k=1$ & (0, 4) & - & 1.429 & 67.8\% & - \\
    \midrule
    \multirow{4}{*}{$k = 1.05$} & (1, 3) & 83 & 1.246 & 67.9\% & 0.1\% \\
    & (2, 2) & 154 & 1.290 & 70.1\% & 2.3\% \\
    & (3, 1) & 205 & 1.372 & 70.6\% & 2.8\% \\
    & (4, 0) & 242 & 1.399 & 70.1\% & 2.3\% \\
    \midrule
    \multirow{4}{*}{$k = 1.1$} & (1, 3) & 38 & 1.193 & 68.5\% & 0.7\% \\
    & (2, 2) & 87 & 1.250 & 70.5\% & 2.7\% \\
    & (3, 1) & 127 & 1.327 & 70.7\% & 2.9\% \\
    & (4, 0) & 160 & 1.415 & 70.5\% & 2.7\% \\
    \bottomrule
\end{tabular}

\begin{comment}
%105
\begin{tabular}{c|c|c|c|c}
    \toprule
    \makecell{$k=1.05$\\$B_L=500$} & $B_S$ & \makecell{testing\\loss} & \makecell{testing\\accuracy} & \makecell{accuracy\\improvement} \\
    \midrule
    $n_S=0, n_L=4$ & - & 1.429 & 67.8\% & - \\
    \midrule
    $n_S=1, n_L=3$ & 83 & 1.246 & 67.9\% & 0.1\% \\
    $n_S=2, n_L=2$ & 154 & 1.290 & 70.1\% & 2.3\% \\
    $n_S=3, n_L=1$ & 205 & 1.372 & 70.6\% & 2.8\% \\
    $n_S=4, n_L=0$ & 242 & 1.399 & 70.1\% & 2.3\% \\
    \bottomrule
\end{tabular}
%110
\begin{tabular}{c|c|c|c|c}
    \toprule
    \makecell{$k=1.1$\\$B_L=500$} & $B_S$ & \makecell{testing\\loss}  & \makecell{testing\\accuracy} & \makecell{accuracy\\improvement} \\
    \midrule
    $n_S=0, n_L=4$ & - & 1.429 & 67.8\% & - \\
    \midrule
    $n_S=1, n_L=3$ & 38 & 1.193 & 68.5\% & 0.7\% \\
    $n_S=2, n_L=2$ & 87 & 1.250 & 70.5\% & 2.7\% \\
    $n_S=3, n_L=1$ & 127 & 1.327 & 70.7\% & 2.9\% \\
    $n_S=4, n_L=0$ & 160 & 1.415 & 70.5\% & 2.7\% \\
    \bottomrule
\end{tabular}
\end{comment}

    }
\end{table}

The experimental results support our claim.
Despite the fact that a worker with a small batch size can achieve low training loss, the number of such workers must be sufficient to increase the final accuracy.
Using only one small-batch worker during training will severely limit the amount of data processed by the small batches.
In this experiment, if we use only one worker with a small batch size, it trains only 21\% of the total data.
Consequently, even though the loss is minimized, the accuracy is still not the best.
In contrast, using three small-batch workers provides 74\% of the entire data for the small batches.
Although the loss is not the lowest, this setting achieves the highest accuracy among all configurations.

\paragraph{\underline{$\mathbf{k=1.1}$}}
Next, to further validate our claim, we conduct another set of experiments with a different extra training time ratio, $k$ = 1.1.
The training parameters are listed in the lower part of Table~\ref{tab:merge_setting}.

Once again, we evaluate the accuracy of Equation~\ref{eq:train_time} for $k$ = 1.1.
The lower part of Table~\ref{tab:time_merge} shows the timing results and relative error across different batch sizes and data amounts.
The results indicate that the maximum relative error is -2.5\%, occurring at a batch size of 38 and a corresponding data amount of 8,750.
Thus, the prediction is sufficiently accurate.

\begin{figure}
    \centering
    \includegraphics[width=\CustomizedWidth]{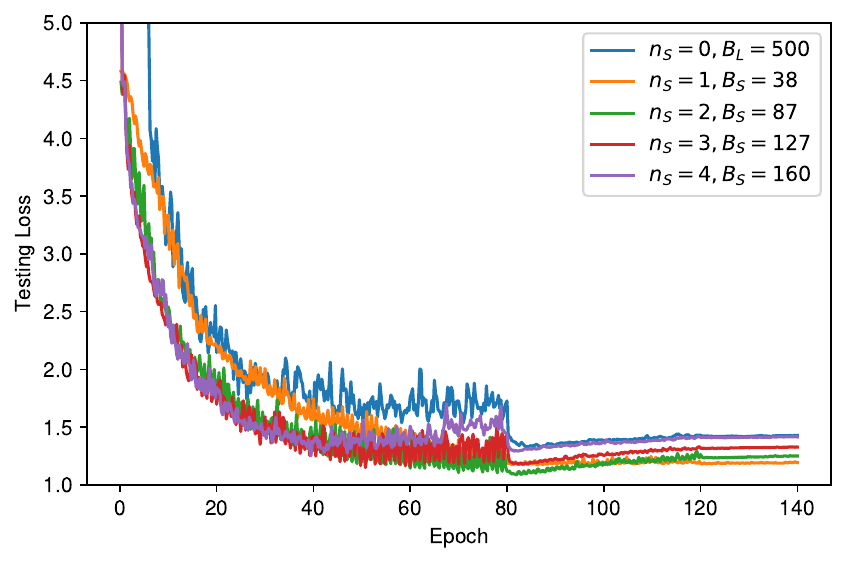}
    \caption{
        Effect of different numbers of small-batch workers on testing loss when $k = 1.1$.
    }
    \label{fig:loss_110}
\end{figure}

\begin{figure}
    \centering
    \includegraphics[width=\CustomizedWidth]{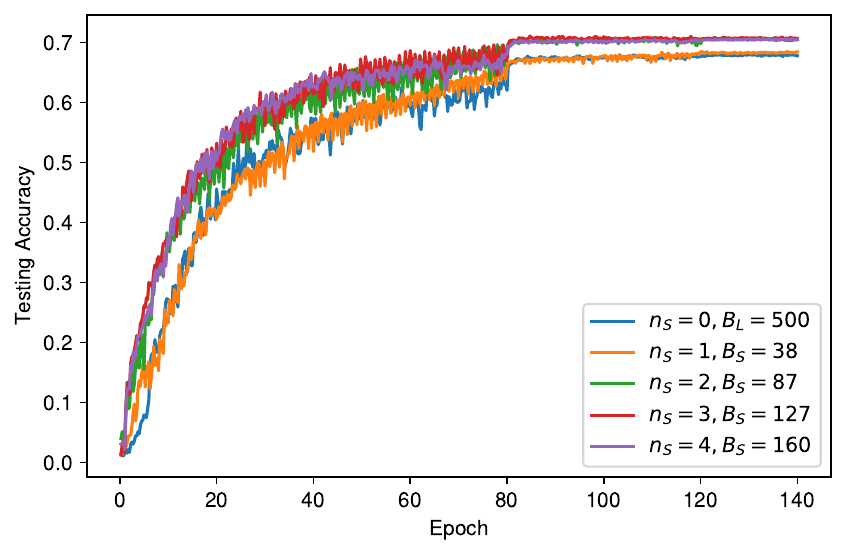} 
    \caption{
        %The testing accuracy of using different numbers of small-batch workers when $k=1.1$.
        Effect of different numbers of small-batch workers on testing accuracy when $k = 1.1$.
    }
    \label{fig:acc_110}
\end{figure}

Figures~\ref{fig:loss_110} and \ref{fig:acc_110} illustrate the testing loss and accuracy during the training process, while the fourth row of Table~\ref{tab:merge_result} provides the final loss and accuracy for all combinations of small and large batch sizes.
We make a similar observation that training with only one small-batch worker achieves the lowest loss but shows only slight accuracy improvement over the baseline as it processes only 18\% of the total data.
In contrast, training with three small-batch workers achieves the highest accuracy of 70.7\%, since these workers process 72\% of the entire dataset.

After conducting these two experiments with different values of the extra training time ratio $k$, we conclude that insufficient small-batch training data results in only marginal accuracy improvement.
Therefore, appropriately increasing the number of small-batch workers in our scheme can help improve accuracy.

\begin{comment}
\subsubsection{Training with Eight Workers}

We extend the previous experiments to eight workers.
Due to the lack of homogeneous devices, all GPU workers are GTX 1080, but the CPUs consist of five Xeon E5-1630 and three i7-5930K, respectively.
The baseline method converges at a loss of 1.342 and an accuracy of 64.5\%.
We set $k$ as 1.05 and 1.1 for the training, and record the experimental results in Table~\ref{tab:8GPU_105} and \ref{tab:8GPU_110}.

\begin{table}
    \centering
    \caption{The training results by using eight workers when $k=1.05$.}
    \label{tab:8GPU_105}
    \adjustbox{max width=\linewidth}{
        \input{tabular/tab_8GPU_105}
    }
\end{table}

\begin{table}
    \centering
    \caption{The training results by using eight workers when $k=1.1$.}
    \label{tab:8GPU_110}
    \adjustbox{max width=\linewidth}{
        \input{tabular/tab_8GPU_110}
    }
\end{table}

We observe similar experimental results from eight workers as from four workers.
We also observe that if we use only one or two small-batch workers, they will see insufficient data, and the training can not converge.
This observation reinforces our claim that the input data for the small-batch workers must be sufficient to improve accuracy. 
The best training results appear when using six or seven small-batch workers, and the amount of data for small-batch training is quite sufficient, accounting for about 74\% to 86\%.
Our approach obtains up to 4.3\% accuracy improvement when $k=1.05$ and up to 4.9\% accuracy improvement when $k=1.1$ over the baseline.
\end{comment}

\subsection{Training with Hybrid Scheme}
\label{sec:exp_HDBL}

%In this section, we compare the performance of the hybrid and dual-batch learning schemes.

\subsubsection{Experimental Setup}

\revisedB{%We use ResNet-18 as the test model and update the training framework to TensorFlow version 2.13.
%In this version, the measured maximum batch size on the GTX 1080 GPU reaches 560 for 32$\times$32 images.
In this experiment, we train ResNet-18 using an updated version of TensorFlow 2.13.
This TensorFlow version provides improved GPU memory utilization and allows a maximum batch size of 560 for 32$\times$32 images on the GTX 1080 GPU.
Therefore, we adopt this updated maximum batch size and re-evaluate all baseline and comparison configurations under the same environment.}
In addition to training on the CIFAR-100 dataset, we also evaluate performance on ImageNet~\cite{russakovsky2015imagenet}, a large-scale image classification dataset containing 1,000 classes, approximately 1.3 million training images, and 50,000 validation images.
We set the model update factor to $\frac{d_S}{d_L}$ and the extra training time ratio to 1.05.

\revisedB{
For the CIFAR-100 experiments, training is conducted using a parameter server framework composed of one server and four workers.
Each worker node is equipped with an Intel Xeon E5-1630 v3 CPU, 64 GB of RAM, and an NVIDIA GeForce GTX 1080 GPU.
The model is trained for a total of 140 epochs.

For the ImageNet experiments, to meet the increased memory and computational demands, training is conducted using a more powerful cluster consisting of one server and four worker nodes.
Each worker node is equipped with an Intel Core i9-12900K CPU, 128 GB of RAM, and one NVIDIA RTX 3090 GPU.
The model is trained for a total of 105 epochs.
To further increase the maximum batch size, we employ mixed-precision training~\cite{micikevicius2018mixed}, an optimization that uses 16-bit floating-point arithmetic instead of 32-bit during training.
}

Table~\ref{tab:J_batch_data} lists the training parameters for CIFAR-100 and ImageNet, derived using the formulas in Section~\ref{sec:DBL}.
For the hybrid scheme, the training process is divided into multiple stages and sub-stages.
As an example, when $n_S = 1$ on CIFAR-100, the small batch sizes $B_{S_1}$ and $B_{S_2}$ are 130 and 102 in the two sub-stages, respectively.
In the following experiments, we use \emph{hybrid} and \emph{DBL} to denote the hybrid and dual-batch learning schemes, respectively.

\begin{table}
    \centering
    \caption{
        Batch size and data amount for CIFAR-100 and ImageNet training with four workers.
    }
    \label{tab:J_batch_data}
    \adjustbox{max width=\linewidth}{
        \begin{tabular}{ccccc}
    \toprule
    CIFAR-100 & \multicolumn{4}{c}{image resolution = (24, 32)} \\
    $k=1.05$ & $B_S$ & $B_L$ & $d_S$ & $d_L$ \\
    \midrule
    $n_S=0$ & - & \multirow{4}{*}{(600, 560)} & - & 12500 \\
    $n_S=1$ & (130, 102) && 10625 & \multirow{3}{*}{13125} \\
    $n_S=2$ & (230, 185) && 11875 & \\
    $n_S=3$ & (294, 243) && 12292 & \\
    $n_S=4$ & (340, 286) & - & 12500 & - \\
    \midrule
    \midrule
    ImageNet & \multicolumn{4}{c}{image resolution = (160, 224, 288)} \\
    $k=1.05$ & $B_S$ & $B_L$ & $d_S$ & $d_L$ \\
    \midrule
    $n_S=0$ & - & \multirow{4}{*}{(2330, 1110, 740)} & - & 320292 \\
    $n_S=1$ & (156, 112, 67) && 272249 & \multirow{3}{*}{336306} \\
    $n_S=2$ & (322, 222, 135) && 304278 & \\
    $n_S=3$ & (464, 310, 190) && 314954 & \\
    $n_S=4$ & (587, 383, 236) & - & 320292 & - \\
    \bottomrule
\end{tabular}
    }
\end{table}

\begin{figure}
    \centering
    \includegraphics[width=\CustomizedWidth]{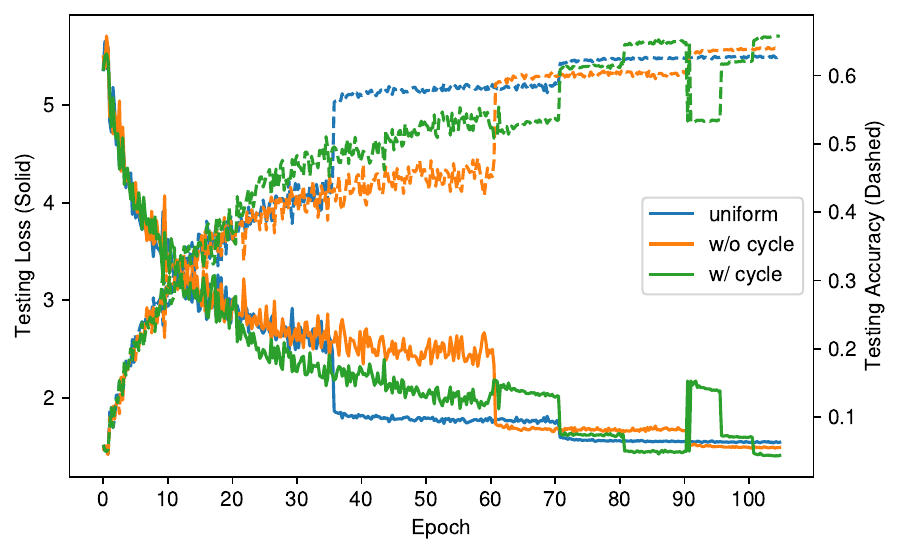}
    \caption{Validation loss and accuracy trajectories comparing Uniform, w/o Cycle, and w/ Cycle schemes.}
    \label{fig:impact_cycle}
\end{figure}

\revisedA{
\subsubsection{Impact of Cyclic Progressive Learning}

To evaluate the effectiveness of the proposed cyclic progressive learning strategy, we disable dual-batch learning by using only the large-batch trainers and compare three training schemes on the ImageNet dataset:
(1) \emph{Uniform}, which evenly divides the training epochs into three stages and decays the learning rate at fixed intervals (i.e., decaying at epochs 35 and 70);
(2) \emph{w/o Cycle}, which uses a heuristic learning rate decay schedule (i.e., decaying at epochs 60 and 90) but trains without cyclic resolution resizing; and
(3) \emph{w/ Cycle}, which uses the same heuristic schedule as scheme two while incorporating the proposed cyclic progressive learning strategy.
Table~\ref{tab:impact_cycle} details the training schedules, final test loss, and test accuracy.
Figure~\ref{fig:impact_cycle} illustrates the validation loss and accuracy curves.

As shown in the results, the proposed \emph{w/ Cycle} scheme achieves the best performance, yielding the lowest loss (1.404) and the highest accuracy (65.8\%).
It outperforms the \emph{w/o Cycle} scheme (64.0\%) and the \emph{Uniform} baseline (62.6\%).
Furthermore, we observe that the accuracy of the \emph{w/ Cycle} scheme at the end of the second stage already exceeds the final accuracy achieved by the other two schemes, and continues to improve during the third stage.
This behavior indicates that revisiting high-resolution images with a refreshed learning rate enables the model to more effectively learn fine-grained features.
These results demonstrate that our cyclic progressive learning strategy is key to enhancing representation learning and model accuracy.
}

\begin{table}
    \centering
    \caption{Comparison of different cyclic schemes on milestones, final loss, and accuracy.}
    \label{tab:impact_cycle}
    \adjustbox{max width=\linewidth}{
        \begin{tabular}{c|ccc}
    \toprule
    Scheme & Uniform & w/o Cycle & w/ Cycle \\
    \midrule
    Milestone & (35, 70) & (60, 90) & (60, 90) \\
    Testing Loss & 1.545 & 1.492 & 1.404 \\
    Testing Accuracy & 62.6\% & 64.0\% & 65.8\% \\
    \bottomrule
\end{tabular}

    }
\end{table}

\subsubsection{CIFAR-100}

For the hybrid scheme on the CIFAR-100 dataset, we split the training into three stages, comprising 80, 40, and 20 epochs, respectively.
In each stage, we evenly divide the training epochs into two sub-stages, which train with image resolutions of 24$\times$24 and 32$\times$32 pixels, respectively.
Table~\ref{tab:J_cifar100_setting} shows the detailed training parameters.

\begin{table}
    \centering
    \caption{
        Training configuration for ResNet-18 on CIFAR-100.
    }
    \label{tab:J_cifar100_setting}
    \adjustbox{max width=\linewidth}{
        \begin{tabular}{cccccccc}
    \toprule
    \multicolumn{2}{c}{stage} & \multicolumn{2}{c}{1} & \multicolumn{2}{c}{2} & \multicolumn{2}{c}{3} \\
    \cmidrule(lr){3-4} \cmidrule(lr){5-6} \cmidrule(lr){7-8}
    \multicolumn{2}{c}{sub-stage} & 1 & 2 & 1 & 2 & 1 & 2 \\
    \midrule
    \multirow{6}{*}{\makecell{dual-batch\\learning}} & epoch & \multicolumn{2}{c}{1-80} & \multicolumn{2}{c}{81-120} & \multicolumn{2}{c}{121-140} \\
    & LR & \multicolumn{2}{c}{0.2} & \multicolumn{2}{c}{0.02} & \multicolumn{2}{c}{0.002} \\
    & $B_S$ & \multicolumn{6}{c}{$B_{S_2}$} \\
    & $B_L$ & \multicolumn{6}{c}{560} \\
    & resolution & \multicolumn{6}{c}{32} \\
    & dropout & \multicolumn{6}{c}{0.2} \\
    \midrule
    \multirow{6}{*}{\makecell{hybrid\\scheme}} & epoch & 1-40 & 41-80 & 81-100 & 101-120 & 121-130 & 131-140 \\
    & LR & \multicolumn{2}{c}{0.2} & \multicolumn{2}{c}{0.02} & \multicolumn{2}{c}{0.002} \\
    & $B_S$ & $B_{S_1}$ & $B_{S_2}$ & $B_{S_1}$ & $B_{S_2}$ & $B_{S_1}$ & $B_{S_2}$ \\
    & $B_L$ & 600 & 560 & 600 & 560 & 600 & 560 \\
    & resolution & 24 & 32 & 24 & 32 & 24 & 32 \\
    & dropout & 0.1 & 0.2 & 0.1 & 0.2 & 0.1 & 0.2 \\
    \bottomrule
\end{tabular}

\begin{comment}
\begin{tabular}{cc|c|c|c|c|c|c}
    \toprule
    \multicolumn{2}{c|}{stage} & \multicolumn{2}{c|}{1} & \multicolumn{2}{c|}{2} & \multicolumn{2}{c}{3} \\
    \multicolumn{2}{c|}{sub-stage} & 1 & 2 & 1 & 2 & 1 & 2 \\
    \midrule
    \multirow{6}{*}{\makecell{dual-batch\\learning}} & epoch & \multicolumn{2}{c|}{1-80} & \multicolumn{2}{c|}{81-120} & \multicolumn{2}{c}{121-140} \\
    & LR & \multicolumn{2}{c|}{0.2} & \multicolumn{2}{c|}{0.02} & \multicolumn{2}{c}{0.002} \\
    & $B_S$ & \multicolumn{6}{c}{$B_{S_2}$} \\
    & $B_L$ & \multicolumn{6}{c}{560} \\
    & resolution & \multicolumn{6}{c}{32} \\
    & dropout & \multicolumn{6}{c}{0.2} \\
    \midrule
    \multirow{6}{*}{\makecell{hybrid\\scheme}} & epoch & 1-40 & 41-80 & 81-100 & 101-120 & 121-130 & 131-140 \\
    & LR & \multicolumn{2}{c|}{0.2} & \multicolumn{2}{c|}{0.02} & \multicolumn{2}{c}{0.002} \\
    & $B_S$ & $B_{S_1}$ & $B_{S_2}$ & $B_{S_1}$ & $B_{S_2}$ & $B_{S_1}$ & $B_{S_2}$ \\
    & $B_L$ & 600 & 560 & 600 & 560 & 600 & 560 \\
    & resolution & 24 & 32 & 24 & 32 & 24 & 32 \\
    & dropout & 0.1 & 0.2 & 0.1 & 0.2 & 0.1 & 0.2 \\
    \bottomrule
\end{tabular}
\end{comment}
    }
\end{table}

\begin{figure}
    \centering
    \includegraphics[width=\CustomizedWidth]{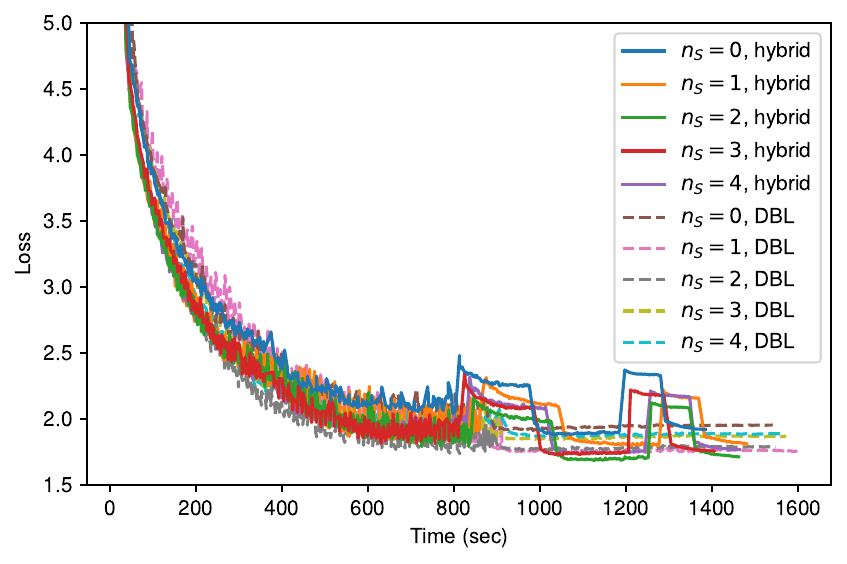} 
    \caption{
        Validation loss for ResNet-18 trained on CIFAR-100 with different learning schemes.
    }
    \label{fig:J_cifar100_loss_105}
\end{figure}

\begin{figure}
    \centering
    \includegraphics[width=\CustomizedWidth]{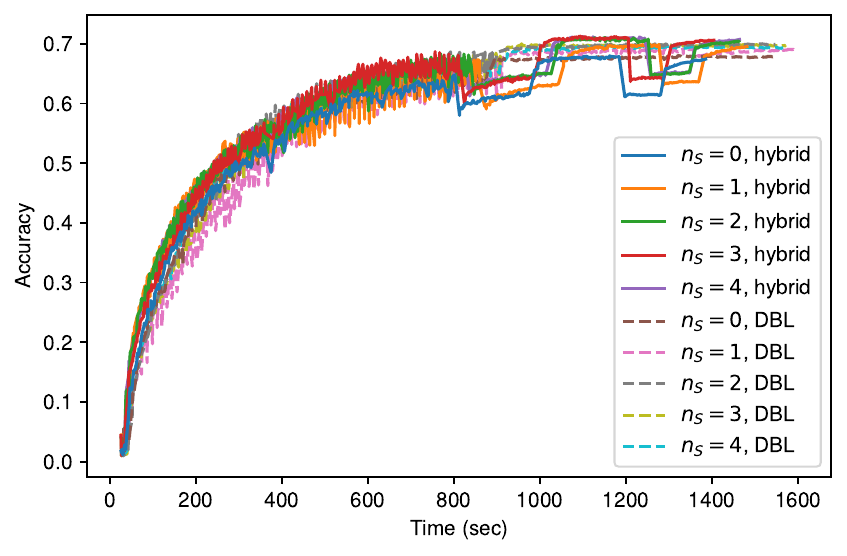} 
    \caption{
        Validation accuracy for ResNet-18 trained on CIFAR-100 with different learning schemes.
    }
    \label{fig:J_cifar100_acc_105}
\end{figure}

Figures~\ref{fig:J_cifar100_loss_105} and \ref{fig:J_cifar100_acc_105} show the testing loss and accuracy for the training process of CIFAR-100, while Table~\ref{tab:J_cifar100_result} lists the final training results for all worker combinations.
The results indicate that, compared to using the dual-batch learning scheme only, the hybrid scheme achieves better testing accuracy in all cases.
Moreover, the hybrid scheme achieves 10.1\% training time reduction, where dual-batch learning takes 1,541 seconds to train the model but the hybrid scheme requires only 1,385 seconds.
The reduction in training time is due to training with lower-resolution images (24$\times$24), and the improvement in accuracy can be attributed to training with images of multiple resolutions, thereby enhancing model generalization.
%Overall, the training time is reduced by 5.5\% when $n_S=2$ and by 10.6\% when $n_S=3$.
%The possible reason for this 5\% gap is that our algorithm determining the training parameters beforehand, whereas the computing speed fluctuates during the actual training process.
%Furthermore, in terms of accuracy, the hybrid scheme outperforms dual-batch learning in all cases.
%We infer that this may be due to the use of images with multiple resolutions, which enhances model generalization and thereby improves accuracy.

\begin{table}
    \centering
    \caption{
        Training results for ResNet-18 on CIFAR-100.
    }
    \label{tab:J_cifar100_result}
    \adjustbox{max width=\linewidth}{
        \begin{tabular}{cccccc}
    \toprule
    \multicolumn{2}{c}{\makecell{$k=1.05$\\4 workers}} & \makecell{training\\time (sec)} & loss & accuracy & \makecell{accuracy\\improvement} \\
    \midrule
    \multirow{5}{*}{\makecell{dual-batch\\learning}} & $n_S = 0$ & 1540.6 & 1.948 & 68.0\% & 0.0\% \\
    \cmidrule{2-6}
    & $n_S = 1$ & 1598.1 & 1.755 & 69.1\% & 1.1\% \\
    & $n_S = 2$ & 1547.1 & 1.787 & 70.0\% & 2.0\% \\
    & $n_S = 3$ & 1569.9 & 1.854 & 70.0\% & 2.0\% \\
    & $n_S = 4$ & 1565.3 & 1.875 & 69.8\% & 1.8\% \\
    \midrule
    \multirow{5}{*}{\makecell{hybrid\\scheme}} & $n_S = 0$ & 1384.7 & 1.893 & 68.0\% & - \\
    \cmidrule{2-6}
    & $n_S = 1$ & 1480.8 & 1.805 & 69.9\% & 1.9\% \\
    & $n_S = 2$ & 1461.7 & 1.691 & 71.0\% & 3.0\% \\
    & $n_S = 3$ & 1404.1 & 1.729 & 71.3\% & 3.3\% \\
    & $n_S = 4$ & 1463.4 & 1.735 & 71.2\% & 3.2\% \\
    \bottomrule
\end{tabular}

\begin{comment}
\begin{tabular}{cc|c|c|c|c}
    \toprule
    \multicolumn{2}{c|}{\makecell{$k=1.05$\\4 workers}} & \makecell{training\\time (sec)} & loss & accuracy & \makecell{accuracy\\improvement} \\
    \midrule
    \multirow{5}{*}{\makecell{dual-batch\\learning}} & $n_S = 0$ & 1540.6 & 1.948 & 68.0\% & 0.0\% \\
    \cmidrule{2-6}
    & $n_S = 1$ & 1598.1 & 1.755 & 69.1\% & 1.1\% \\
    & $n_S = 2$ & 1547.1 & 1.787 & 70.0\% & 2.0\% \\
    & $n_S = 3$ & 1569.9 & 1.854 & 70.0\% & 2.0\% \\
    & $n_S = 4$ & 1565.3 & 1.875 & 69.8\% & 1.8\% \\
    \midrule
    \multirow{5}{*}{\makecell{hybrid\\scheme}} & $n_S = 0$ & 1384.7 & 1.893 & 68.0\% & - \\
    \cmidrule{2-6}
    & $n_S = 1$ & 1480.8 & 1.805 & 69.9\% & 1.9\% \\
    & $n_S = 2$ & 1461.7 & 1.691 & 71.0\% & 3.0\% \\
    & $n_S = 3$ & 1404.1 & 1.729 & 71.3\% & 3.3\% \\
    & $n_S = 4$ & 1463.4 & 1.735 & 71.2\% & 3.2\% \\
    \bottomrule
\end{tabular}
\end{comment}
    }
\end{table}

\subsubsection{ImageNet}
\label{sec:exp_hybrid_imagenet}

For the ImageNet dataset, the training process is divided into three stages, each comprising three sub-stages that handle image resolutions of 160, 224, and 288, respectively.
Table~\ref{tab:J_imagenet_setting} depicts the training parameters.
%, where training is repeated for three different image resolutions while the learning rate gradually decreases during the training process.

\begin{table*}
    \centering
    \caption{
        Training configuration for ResNet-18 on ImageNet.
    }
    \label{tab:J_imagenet_setting}
    \adjustbox{max width=\linewidth}{
        \begin{tabular}{ccccccccccc}
    \toprule
    \multicolumn{2}{c}{stage} & \multicolumn{3}{c}{1} & \multicolumn{3}{c}{2} & \multicolumn{3}{c}{3} \\
    \cmidrule(lr){3-5} \cmidrule(lr){6-8} \cmidrule(lr){9-11}
    \multicolumn{2}{c}{sub-stage} & 1 & 2 & 3 & 1 & 2 & 3 & 1 & 2 & 3 \\
    \midrule
    \multirow{6}{*}{\makecell{dual-batch\\learning}} & epoch & \multicolumn{3}{c}{1-60} & \multicolumn{3}{c}{61-90} & \multicolumn{3}{c}{91-105} \\
    & LR & \multicolumn{3}{c}{0.2} & \multicolumn{3}{c}{0.02} & \multicolumn{3}{c}{0.002} \\
    & $B_S$ & \multicolumn{9}{c}{$B_{S_3}$} \\
    & $B_L$ & \multicolumn{9}{c}{740} \\
    & resolution & \multicolumn{9}{c}{288} \\
    & dropout & \multicolumn{9}{c}{0.3} \\
    \midrule
    \multirow{6}{*}{\makecell{hybrid\\scheme}} & epoch & 1-20 & 21-40 & 41-60 & 61-70 & 71-80 & 81-90 & 91-95 & 96-100 & 101-105 \\
    & LR & \multicolumn{3}{c}{0.2} & \multicolumn{3}{c}{0.02} & \multicolumn{3}{c}{0.002} \\
    & $B_S$ & $B_{S_1}$ & $B_{S_2}$ & $B_{S_3}$ & $B_{S_1}$ & $B_{S_2}$ & $B_{S_3}$ & $B_{S_1}$ & $B_{S_2}$ & $B_{S_3}$ \\
    & $B_L$ & 2330 & 1110 & 740 & 2330 & 1110 & 740 & 2330 & 1110 & 740 \\
    & resolution & 160 & 224 & 288 & 160 & 224 & 288 & 160 & 224 & 288 \\
    & dropout & 0.1 & 0.2 & 0.3 & 0.1 & 0.2 & 0.3 & 0.1 & 0.2 & 0.3 \\
    \bottomrule
\end{tabular}

\begin{comment}
\begin{tabular}{cc|c|c|c|c|c|c|c|c|c}
    \toprule
    \multicolumn{2}{c|}{stage} & \multicolumn{3}{c|}{1} & \multicolumn{3}{c|}{2} & \multicolumn{3}{c}{3} \\
    \multicolumn{2}{c|}{sub-stage} & 1 & 2 & 3 & 1 & 2 & 3 & 1 & 2 & 3 \\
    \midrule
    \multirow{6}{*}{\makecell{dual-batch\\learning}} & epoch & \multicolumn{3}{c|}{1-60} & \multicolumn{3}{c|}{61-90} & \multicolumn{3}{c}{91-105} \\
    & LR & \multicolumn{3}{c|}{0.2} & \multicolumn{3}{c|}{0.02} & \multicolumn{3}{c}{0.002} \\
    & $B_S$ & \multicolumn{9}{c}{$B_{S_3}$} \\
    & $B_L$ & \multicolumn{9}{c}{740} \\
    & resolution & \multicolumn{9}{c}{288} \\
    & dropout & \multicolumn{9}{c}{0.3} \\
    \midrule
    \multirow{6}{*}{\makecell{hybrid\\scheme}} & epoch & 1-20 & 21-40 & 41-60 & 61-70 & 71-80 & 81-90 & 91-95 & 96-100 & 101-105 \\
    & LR & \multicolumn{3}{c|}{0.2} & \multicolumn{3}{c|}{0.02} & \multicolumn{3}{c}{0.002} \\
    & $B_S$ & $B_{S_1}$ & $B_{S_2}$ & $B_{S_3}$ & $B_{S_1}$ & $B_{S_2}$ & $B_{S_3}$ & $B_{S_1}$ & $B_{S_2}$ & $B_{S_3}$ \\
    & $B_L$ & 2330 & 1110 & 740 & 2330 & 1110 & 740 & 2330 & 1110 & 740 \\
    & resolution & 160 & 224 & 288 & 160 & 224 & 288 & 160 & 224 & 288 \\
    & dropout & 0.1 & 0.2 & 0.3 & 0.1 & 0.2 & 0.3 & 0.1 & 0.2 & 0.3 \\
    \bottomrule
\end{tabular}
\end{comment}
    }
\end{table*}

Table~\ref{tab:J_imagenet_result} shows the final training results.
As observed from the results, the accuracy improves as more small-batch workers are introduced with both the dual-batch learning and hybrid methods.
In addition, when the number of small-batch workers is not sufficient (e.g., less than or equal to half), the accuracy is lower than the baseline method (all large-batch workers).
However, when we increase the number of small-batch workers to more than two, both methods achieve similar performance to the baseline.

\begin{table}
    \centering
    \caption{
        Training results for ResNet-18 on ImageNet.
    }
    \label{tab:J_imagenet_result}
    \adjustbox{max width=\linewidth}{
        \begin{tabular}{cccccc}
    \toprule
    \multicolumn{2}{c}{\makecell{$k=1.05$\\4 workers}} & \makecell{training\\time (sec)} & loss & accuracy & \makecell{accuracy\\improvement} \\
    \midrule
    \multirow{5}{*}{\makecell{dual-batch\\learning}} & $n_S = 0$ & 33974.6 & 1.874 & 65.9\% & 0.8\% \\
    \cmidrule{2-6}
    & $n_S = 1$ & 34454.0 & 2.226 & 62.1\% & -3.0\% \\
    & $n_S = 2$ & 34203.7 & 2.032 & 64.5\% & -0.6\% \\
    & $n_S = 3$ & 33997.3 & 1.926 & 65.7\% & 0.6\% \\
    & $n_S = 4$ & 33917.9 & 1.913 & 65.5\% & 0.4\% \\
    \midrule
    \multirow{5}{*}{\makecell{hybrid\\scheme}} & $n_S = 0$ & 21757.8 & 1.949 & 65.1\% & - \\
    \cmidrule{2-6}
    & $n_S = 1$ & 22160.9 & 2.138 & 62.8\% & -2.3\% \\
    & $n_S = 2$ & 22277.6 & 1.971 & 64.6\% & -0.5\% \\
    & $n_S = 3$ & 22245.4 & 1.936 & 65.2\% & 0.1\% \\
    & $n_S = 4$ & 22162.8 & 1.884 & 65.5\% & 0.4\% \\
    \bottomrule
\end{tabular}

\begin{comment}
\begin{tabular}{cc|c|c|c|c}
    \toprule
    \multicolumn{2}{c|}{\makecell{$k=1.05$\\4 workers}} & \makecell{training\\time (sec)} & loss & accuracy & \makecell{accuracy\\improvement} \\
    \midrule
    \multirow{5}{*}{\makecell{dual-batch\\learning}} & $n_S = 0$ & 33974.6 & 1.874 & 65.9\% & 0.8\% \\
    \cmidrule{2-6}
    & $n_S = 1$ & 34454.0 & 2.226 & 62.1\% & -3.0\% \\
    & $n_S = 2$ & 34203.7 & 2.032 & 64.5\% & -0.6\% \\
    & $n_S = 3$ & 33997.3 & 1.926 & 65.7\% & 0.6\% \\
    & $n_S = 4$ & 33917.9 & 1.913 & 65.5\% & 0.4\% \\
    \midrule
    \multirow{5}{*}{\makecell{hybrid\\scheme}} & $n_S = 0$ & 21757.8 & 1.949 & 65.1\% & - \\
    \cmidrule{2-6}
    & $n_S = 1$ & 22160.9 & 2.138 & 62.8\% & -2.3\% \\
    & $n_S = 2$ & 22277.6 & 1.971 & 64.6\% & -0.5\% \\
    & $n_S = 3$ & 22245.4 & 1.936 & 65.2\% & 0.1\% \\
    & $n_S = 4$ & 22162.8 & 1.884 & 65.5\% & 0.4\% \\
    \bottomrule
\end{tabular}
\end{comment}
    }
\end{table}

\revisedB{On the other hand, the training efficiency is substantially enhanced under the hybrid scheme.
Across all matched configurations from $n_S = 0$ to $n_S = 4$ in Table~\ref{tab:J_imagenet_result}, the hybrid scheme consistently reduces training time by 34.6\% to 36.0\% compared to dual-batch learning.
For example, with $n_S = 4$, the training time decreases from 33,918 seconds to 22,163 seconds, achieving a 34.7\% reduction while maintaining an identical validation accuracy of 65.5\%.
Furthermore, across all configurations, the validation accuracy differences between the hybrid scheme and dual-batch learning remain within statistical fluctuations of 0.8\%, demonstrating that the acceleration is achieved without severe accuracy degradation.
%In terms of time-to-accuracy, the hybrid scheme reaches 65.5\% accuracy 3.2 hours faster than dual-batch learning, yielding a 1.53$\times$ speedup.
In terms of time-to-accuracy, the best-performing hybrid scheme (65.5\% accuracy) reduces training time by 3.2 hours compared with dual-batch learning only, corresponding to a 1.53$\times$ speedup.
We also observe that the hybrid scheme achieves a significant training time reduction on ImageNet, compared to 10.1\% on CIFAR-100.}
This is due to the size ratio between the small-resolution and large-resolution images.
The size ratio is 0.31 ($160^2/288^2$) on ImageNet but is 0.56 ($24^2/32^2$) on CIFAR-100.
As a result, the hybrid scheme trains the model using relatively smaller images from the training dataset, thus leading to better training efficiency.

\revisedB{
\subsection{Scalability and Multi-Worker Speedup Analysis}
\label{sec:exp_scalability}

To evaluate the performance benefits and scalability of distributed training, we investigate the hybrid scheme across varying numbers of worker nodes using ResNet-18 on the ImageNet dataset.
Figure~\ref{fig:scalability} illustrates the validation accuracy over training time for configurations ranging from a single worker to four workers.

\begin{figure}
    \centering
    \includegraphics[width=\CustomizedWidth]{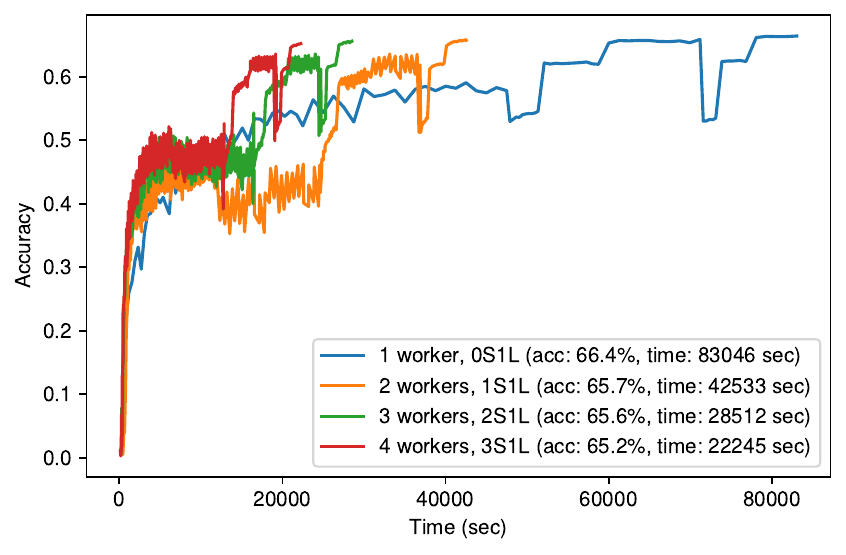}
    \caption{Validation accuracy over training time for ResNet-18 on ImageNet under different numbers of workers.}
    \label{fig:scalability}
\end{figure}

Specifically, for $n=1$, the single-worker baseline adopts the standard large-batch configuration denoted as $0S1L$, where $S$ and $L$ represent the numbers of small-batch and large-batch workers, respectively.
For multi-worker setups with $n$ from 2 to 4, we deploy 1 to 3 small-batch workers, pairing a single large-batch worker with $n-1$ small-batch workers.
This configuration choice is guided by our empirical findings in Sections~\ref{sec:exp_DBL} and~\ref{sec:exp_HDBL}.
Allocating the majority of workers to small-batch training guarantees stable loss reduction and prevents severe accuracy degradation.
Consequently, this strategy preserves model generalization while capitalizing on the high throughput of large-batch processing.

As shown in Figure~\ref{fig:scalability}, distributed training yields substantial speedups compared to single-GPU training.
Training ResNet-18 for 105 epochs on a single RTX 3090 GPU requires 23.1 hours, which poses a significant bottleneck for iterative model development and research.
By scaling to 2, 3, and 4 workers, the overall training time decreases to 11.8, 7.9, and 6.2 hours, respectively.
This corresponds to speedups of $1.95\times$, $2.91\times$, and $3.73\times$ relative to the single-GPU baseline, achieving near-linear scaling efficiency.
Nevertheless, the system still reflects the inherent challenges of asynchronous distributed training.
As more workers participate, the final validation accuracy slightly decreases from 66.4\% to 65.2\%, indicating increased gradient staleness and difficulty of reaching parameter consensus across worker nodes.
These results demonstrate that the proposed hybrid scheme effectively utilizes distributed cluster resources to reduce training time while keeping accuracy degradation bounded within an acceptable 1.2\% margin.
}

\revisedA{
\subsection{Systems and Communication Overhead Analysis}
\label{sec:systems_analysis}

Communication overhead is a critical factor influencing scalability and overall training efficiency for distributed training systems based on the parameter server.
%To evaluate the systems-level performance of our proposed schemes, we 
This section provides a qualitative analysis of the communication overhead, communication-computation overlap, and network utilization under the hybrid dual-batch and cyclic progressive learning scheme.

First, the proposed \emph{model-update factor} in dual-batch learning incurs \emph{no additional communication overhead}.
The model-update factor is applied locally at each worker node by scaling the locally computed gradients before transmitting them to the parameter server.
Hence, the size of data transmitted over the network (e.g., parameters and gradients) remains identical to the standard ASP-SGD, without introducing additional communication traffic or network protocol overhead.

Second, our asynchronous training strategy naturally overlaps communication with computation.
Each worker independently pulls model parameters from the parameter server, performs local training, and pushes gradient updates without global synchronization barriers.
%This asynchronous update model enables communication and computation to be highly overlapped in time.
As a result, parameter communication can proceed concurrently with GPU computation, effectively hiding the communication latency.
This asynchronous execution not only improves scalability but also avoids the straggler problem commonly encountered in BSP training.

Third, our CPL scheme further reduces network traffic during the early training phases.
%In conventional progressive learning, decreasing the image resolution shortens the computation time per step, which increases the frequency of network updates and can bottleneck the network bandwidth (the communication-to-computation ratio increases).
%However, CPL addresses this bottleneck by employing an \emph{adaptive batch size}.
During training phases with low-resolution images, CPL adaptively increases the batch size, thereby reducing the number of update steps (i.e., $\lceil d / B \rceil$) and the communication overhead required to process one epoch.
This is in contrast to conventional progressive learning, which uses a fixed batch size throughout training and maintains the same communication frequency regardless of the input image resolutions.
By reducing the frequency of communication, our adaptive batch size strategy alleviates network bandwidth pressure and maintains high distributed training throughput across different resolution phases.
}

\subsection{Selection of Maximum Batch Size}

Selecting an appropriate batch size is crucial in deep learning training, as it directly impacts GPU utilization, training efficiency, and overall performance.
Our distributed training schemes begin by selecting the large-batch size ($B_L$), and then determine the small-batch size ($B_S$) through the configuration of the number of small- and large-batch workers ($n_S$ and $n_L$), based on Equations~\ref{eq:sametime1} to~\ref{eq:find_BS}.
In these schemes, the large-batch size ($B_L$) should be as large as possible to fully utilize GPU resources and minimize training time.
Hence, the key question is how to determine $B_L$ before applying our schemes.
A typical approach employed by most practitioners is through trial and error, gradually increasing the batch size until the training frameworks (e.g., PyTorch and TensorFlow) trigger a GPU out-of-memory error.
To address this issue, we present a method to automate finding $B_L$ based on linear regression by profiling memory usage, without introducing significant computational overhead.

%For most tasks, the maximum batch size $B_{max}$ requires manual input from the user.
%In this section, we explore a preliminary approach to automating this parameter.
%To address this, we propose integrating a program to calculate $B_{max}$ before training begins.
%As is well known, machine learning frameworks (e.g., PyTorch, TensorFlow, etc.) will crash if the GPU cannot allocate memory beyond its available free memory.
%In practice, users often determine $B_{max}$ through trial and error, which is also the approach adopted in this paper.
%We believe there is a feasible way to automate $B_{max}$ by profiling memory usage without introducing significant additional computational overhead.

Assuming that the memory usage scales linearly with batch size $B$, we can estimate the total memory usage $M(B)$ required for training using Equation~\ref{eq:mem_usage}:
\begin{equation}
    M(B) = \sum_{l=1}^L p_l + B \times \sum_{l=1}^L a_l     \label{eq:mem_usage}
\end{equation}
where $p_l$ represents the memory required for the parameters (weights, gradients, etc.) of the $l$-th layer, $a_l$ denotes the memory required for storing and computing the activations per sample at the $l$-th layer, and $L$ is the total number of layers.

Our goal is to determine the maximum batch size $B_{max}$ such that $M(B_{max}) \leq M_{GPU}$, where $M_{GPU}$ is the total GPU memory size.
To estimate this, we first run several batches of varying sizes, measure their memory consumption, and construct a memory usage model $M(B)$ via linear regression.
This model then enables the estimation of the maximum feasible batch size $B_{max}$ for the given GPU memory $M_{GPU}$.

\begin{figure}
    \centering
    \includegraphics[width=\CustomizedWidth]{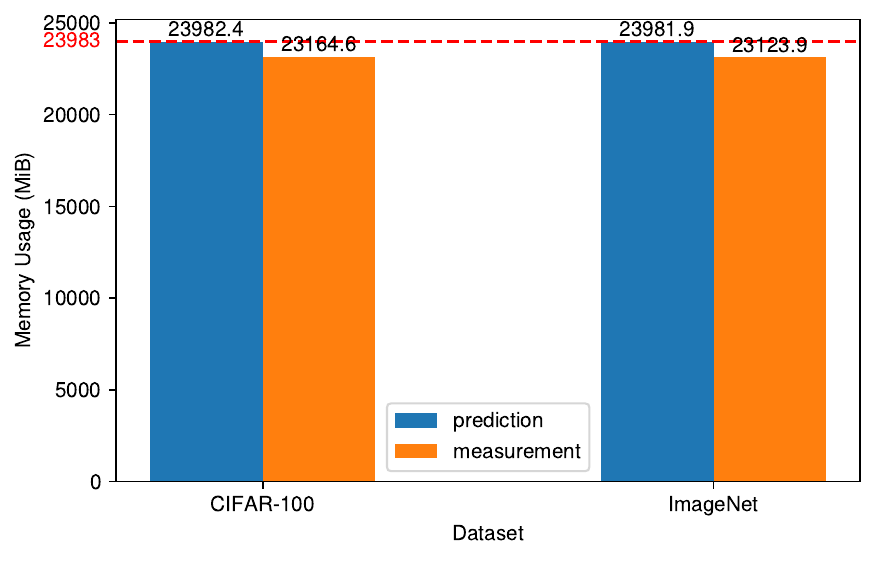} 
    \caption{
        Comparison of the measured and predicted memory usage for training ResNet-18 on an NVIDIA RTX 3090 GPU.
        The predicted maximum batch size is 11,147 for image resolution of 32$\times$32 on CIFAR-100 and 1,345 for image resolution of 224$\times$224 on ImageNet.
    }
    \label{fig:mem_usage_reg}
\end{figure}

To verify the prediction accuracy of the derived memory usage model, we conduct experiments on the NVIDIA RTX 3090 GPU with 24 GB of on-device memory, and select eight batches of different sizes (64, 128, 192, 256, 320, 384, 448, 512) for linear regression.
Figure~\ref{fig:mem_usage_reg} shows the measured and predicted maximum memory usage for the CIFAR-100 and ImageNet datasets.
For ImageNet, we use mixed precision training, as in Section~\ref{sec:exp_hybrid_imagenet}.
%We conduct an experiment to validate this approach, and the results of memory usage are shown in Figure~\ref{fig:mem_usage_gap}.
%We select PyTorch version 2.5, using an NVIDIA RTX 3090 GPU with 24 GB of memory, as the test platform.
We obtain maximum batch sizes of 11,147 for CIFAR-100 with an image resolution of 32$\times$32 and 1,345 for ImageNet with an image resolution of 224$\times$224.
The corresponding relative error between the predicted and measured values is approximately 3.5\% on CIFAR-100 and 3.7\% on ImageNet.
%The slight discrepancy between the predicted and measured values results from some memory usage reserved by the training framework and the cuDNN library.
%Since the measured values are consistently lower than the predicted ones, the computed batch size $B_{comp}$ is smaller than the actual maximum batch size $B_{max}$.
%Although $B_{comp}$ is not optimal, it remains a sub-optimal yet practical solution.
We conclude that this method is feasible and reliable for our dual-batch learning and hybrid learning schemes.
%Moreover, since this is a preliminary idea with coarse profiling, there remains room for optimization through finer profiling in future work.

\section{Discussion and Limitations}
\label{sec:limitations}

\revisedA{
\subsection{Orthogonality to Communication Compression}

Recent advances in distributed deep learning have introduced various communication-efficient optimization techniques, including gradient quantization, sparsification, and compression algorithms such as Deep Gradient Compression (DGC)~\cite{lin2018deep} and PowerSGD~\cite{vogels2019powersgd}.
These methods focus on reducing the amount of transmitted gradient data to alleviate communication bottlenecks in large-scale distributed training.
However, these compression techniques primarily target communication efficiency and typically employ a single and fixed large batch size across all workers.
They do not address the generalization degradation caused by large-batch training and consequently still suffer from lower model accuracy on test datasets.

In contrast, our dual-batch learning scheme overcomes the generalization degradation of large-batch training and is complementary to these communication-efficient methods.

\subsection{Homogeneous Cluster Assumption}

Our training time prediction model (Equation~\ref{eq:train_time}) is designed for homogeneous distributed training environments, where all worker nodes have identical GPU hardware and network configurations.
%In heterogeneous environments with mixed GPU architectures or non-uniform network bandwidth, the profiling coefficients ($a$ and $b$) would differ across workers, making the performance model more complex.
In heterogeneous environments with mixed GPU architectures or non-uniform network bandwidth, the profiling coefficients ($a$ and $b$) would differ across workers.
This variation requires worker-specific profiling parameters and significantly increases modeling complexity.
Therefore, extending the proposed load-balancing and adaptive batch size framework to heterogeneous distributed systems is beyond the scope of this work and remains an interesting direction for future work.
}

\subsection{Vision Transformer Incompatibility}

Our proposed scheme is designed for convolutional neural networks (CNNs) and cannot be directly applied to Vision Transformer (ViT)-based models, as the ViT architecture fundamentally differs from that of CNNs.
In CNNs, input images can have varying resolutions, and the use of a global pooling layer ensures outputs of consistent dimensions.
This flexibility allows our hybrid scheme to accelerate training through resolution adjustments and larger batch sizes to fully utilize GPU resources.
In contrast, ViTs require fixed input resolutions, as images are divided into a predefined number of patches with fixed spatial dimensions that are linearly projected and processed as tokens.
Altering the input resolution would disrupt the number of patches and the corresponding positional encodings, thus preventing the direct application of our hybrid scheme to ViT architectures.

\section{Conclusion}
\label{sec:conclusion}

In this paper, we introduce \emph{dual-batch learning}, a novel scheme that co-trains CNN models using two different batch sizes.
By allowing a small increase in training time, we achieve improved model generalization compared to using a single batch size.
To further reduce the training time, we propose \emph{cyclic progressive learning}, a scheme that schedules the training process by adjusting image resolution and regularization.
Combining these two methods, we present the \emph{hybrid scheme}, which reduces training time while maintaining strong generalization ability for CNNs by dynamically adjusting both batch sizes and image resolutions during training.
The evaluation demonstrates that our hybrid scheme achieves significant time reduction and comparable accuracy to using dual-batch learning or cyclic progressive learning alone.

%% bibliography
\bibliographystyle{elsarticle-num-names}
\bibliography{references}

\end{document}